\documentclass[letterpaper]{article} 
\usepackage[T1]{fontenc}

\usepackage{geometry}
\usepackage{setspace}

\usepackage[style = chem-acs]{biblatex}
\usepackage{graphicx}
\usepackage{float}
\usepackage{amssymb}
\usepackage{subcaption}
\usepackage{comment}
\newfloat{scheme}{htbp}{los}
\floatname{scheme}{Scheme}
\floatname{chart}{Chart}
\newfloat{graph}{htbp}{loh}
\usepackage{booktabs}
\usepackage{siunitx}
\usepackage{amsmath}

\usepackage{chemformula} 
\usepackage{soul} 
\usepackage[version = 4]{mhchem} 

\usepackage{authblk}
\author[1]{Ioannis M. Koutzoglou*}
\author[1]{Stamatios Amanatiadis}
\affil[1]{School of Electrical and Computer Engineering, Aristotle University of Thessaloniki,
54124 Thessaloniki, Greece}
\author[1]{Nikolaos V. Kantartzis}
\author[2,3]{Theodosios D. Karamanos}
\affil[2]{Laboratoire GeePs, Sorbonne Université, CNRS, 75005 Paris, France}
\affil[3]{Laboratoire GeePs, Université Paris-Saclay, CentraleSupélec, CNRS, 91190 Gif-sur-Yvette, France}

\title{Generalized Conductivity Modeling and Selective Harmonic Amplification in Time-Modulated Graphene Cavities}

\date{*Email: ikoutzo@ece.auth.gr}

\begin{document}

\maketitle

\begin{abstract}
The selective harmonic enhancement in cavities formed by stacks of time-modulated graphene sheets and a reflecting boundary is investigated. A semi-analytic framework based on an operator formulation and the transfer matrix method is developed and validated against a modified finite-difference time-domain algorithm. The temporal dispersion of graphene is treated through both a generalized Taylor-expanded conductivity model and a reduced high-bias approximation. By employing particle swarm optimization to tune the cavity gaps, selected Floquet harmonics are engineered under distinct modulation regimes. Numerical results show strong enhancement of first-order sidebands in the high-bias regime, controlled third-order harmonic generation beyond the linear regime with an explicit trade-off between target amplification and total non-target leakage, and symmetry-induced purely even harmonic generation under zero-centered modulation.
\end{abstract}

\section*{Keywords}
\emph{time-varying media, graphene, dispersive electromagnetic interface, Floquet harmonic amplification, full-wave numerical simulation}



\section{1. Introduction}
Time-varying materials have attracted intense interest in recent years as a powerful platform for manipulating electromagnetic waves beyond the limitations of stationary media~\cite{Koutzoglou2026,galiffi2022photonics, caloz2019spacetime}. By deliberately modulating material properties in time, one introduces an additional degree of freedom that enables phenomena such as frequency translation~\cite{Liu2016ACSPhotonics, Zhou2020}, nonreciprocity~\cite{CorreasSerrano2016, Wang2020, Yu2009, Shi2017ACSPhotonics}, parametric amplification~\cite{asadchy2022parametric, Lyubarov2022, Lee2021, wang2025expanding}, dynamic impedance matching~\cite{PachecoPena2020Optica, Li2021, Castaldi2021}, beam control~\cite{PachecoPena2020LSA}, and temporal cloaking~\cite{Ramaccia2017PRB, Fridman2012Nature, Zhou2019NatCommun}. A distinctive feature of periodically time-modulated media is the generation and coupling of harmonic frequencies~\cite{Wu2020TAP, Taravati2021PRApplied,caloz2019spacetime}, which makes them particularly appealing for compact and reconfigurable frequency-conversion devices across the microwave~\cite{Taravati2022ACSPhotonics}, millimeter-wave~\cite{Peng2025ACSAEM}, and terahertz spectra~\cite{Galiffi2022AdvPhoton}. Over the past years, with the expansion of telecommunication systems beyond $5$G standards, the THz spectrum has become an increasingly appealing candidate. Thus, new efficient setups are required to act as frequency sources, and time-varying media applications are being recognized as increasingly well-suited for this purpose. Additionally, the introduction of time-varying media in optics and photonics is expected to improve the efficiency of the respective applications~\cite{galiffi2022photonics,asgari2024theory}.

In this context, graphene constitutes an especially attractive platform for time-varying electromagnetic systems~\cite{Drahushuk2016ACSNano, Salary2018PRB}. Here, its surface conductivity depends on the chemical potential, which is electrically tunable and can be modulated at high speeds. The restriction, though is that a relatively small bias voltage change is applied to ensure linear dependence between the chemical potential and the surface conductivity. In any case, these properties allow deep temporal modulation and strong light-matter interaction, enabling efficient harmonic generation and frequency mixing in the far-infrared regime~\cite{SensaleRodriguez2012NatComm, altares2017frequency, Sedeh2022TAP}. In particular, graphene is represented as an equivalent surface conductivity that follows the Drude distribution with a time-modulated bias voltage, while the same modeling idea with graphene ribbons was presented as a concept of a MIMO communication system~\cite{Sedeh2022TAP}. When graphene is embedded in resonant configurations, the interaction between temporal modulation and multiple reflections can be exploited to selectively enhance specific modulation sidebands, offering a route toward compact and efficient frequency-agile sources. Specifically, the deployment of time-modulated graphene on the aperture of a slot antenna could create the concept of a nonlinear frequency generator~\cite{amanatiadis2025enhanced}. Moreover, the incorporation of graphene sheets on Fabry--Perot-like structures~\cite{Karamanos2024META} or the assortment of stacked surfaces exhibiting Drude dispersion~\cite{movahediqomi2026stacked} are shown to provide interesting applications for THz and optical regimes. These applications include the enhancement or the minimization of generated harmonics, that are also assisted by physics-based optimization using a common distance between the surfaces as a single degree of freedom. Nevertheless, all these works assume the linear dependence between graphene conductivity and the time-modulated electric bias, which limits the chemical potential bounds. The optimization of graphene sheet arrangements is promising for innovative future devices in the spectra of mm-wave, THz, and optics while a complete modeling of the time modulation of the surface conductivity is required.

In this work, we investigate the controlled enhancement of selected harmonics using a stack of time-modulated graphene sheets terminated by a perfect electric conductor (PEC). A single time-varying graphene sheet already produces a comb of generated harmonics when illuminated by a monochromatic wave. However, arranging multiple such sheets in a cavity configuration provides additional degrees of freedom through multiple scattering and interference, enabling substantially stronger and more selective harmonic amplification. To rigorously describe this problem, we develop an operator-based frequency-domain formulation for dispersive, time-modulated conductive interfaces and specialize it to graphene with Drude-type dispersion. For this reason, a Taylor-expanded scheme is derived for the surface conductivity of graphene, which provides a general and physically transparent description of temporal modulation. This representation accurately captures higher-order harmonic generation, including regimes where the relationship between the surface conductivity and the chemical potential is strongly nonlinear, surpassing the limitation of state-of-the-art works. Then, the conductivity model is embedded into a transfer-matrix framework for multilayer structures, with an equivalent recursive formulation retained for physical insight. The proposed approach is validated against a properly modified finite-difference time-domain (FDTD) solver for dispersive, time-varying conductive surfaces demonstrating excellent agreement across a wide range of bias conditions. Building on this framework, we then address the inverse design problem of harmonic engineering. Using particle swarm optimization (PSO), we show how cavity spacings can be tailored to selectively enhance either first-order sidebands in the high-bias regime or higher-order harmonics in regimes where nonlinear temporal mixing is essential. In the latter case, a multi-objective optimization strategy is employed to balance target enhancement against the suppression of non-target spectral leakage, while a zero-centered modulation regime is shown to produce purely even harmonic generation by symmetry. More generally, the results highlight that allowing nonuniform cavity spacings provides additional design freedom and enables more effective harmonic engineering compared to single-spacing designs. Overall, this work establishes a rigorous and versatile modeling and design framework for dispersive time-modulated graphene cavities and demonstrates their potential as compact, frequency-agile building blocks for next-generation terahertz and mm-wave systems.

\section{2. Electromagnetic Modeling and Validation of Time-Modulated Graphene Structures}

\subsection{2.1 Operator Formulation for a Dispersive Time-Modulated Graphene Sheet}\label{section2.1}

We begin by establishing the electromagnetic response of a single graphene sheet whose surface conductivity is both temporally modulated and dispersive. This elementary configuration constitutes the building block for all subsequent multilayer and resonant structures considered in this work.

\begin{figure*}
    \centering
    \includegraphics[width=0.35\linewidth]{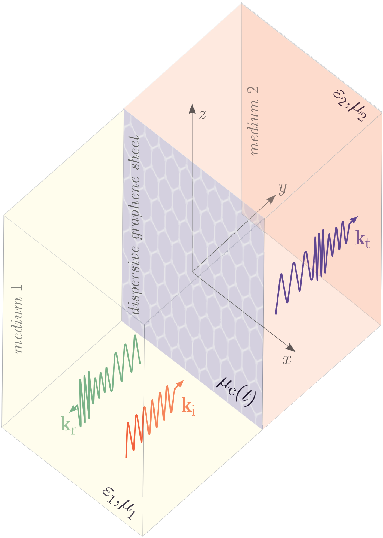}
    \caption{Normal incidence on a single time-varying and dispersive graphene sheet placed at the $xz$-plane.}
    \label{fig1}
\end{figure*}

Let us assume a graphene sheet located in the $xz$-plane and illuminated under normal incidence by a monochromatic plane wave $\mathbf{E}_{\rm i} = E_{\rm i} \cos(\omega t - k_2y)\,\hat{\mathbf{z}},$
where $\omega$ is the angular frequency, $k_2=\omega\sqrt{\mu_2\varepsilon_2}$ is the wavenumber of the surrounding medium, and $\mu_2$, $\varepsilon_2$ denote its constitutive parameters, as illustrated in Figure~\ref{fig1}. For simplicity, the graphene sheet is modeled as an isotropic conductive interface, so its response is fully described by the scalar surface conductivity $\sigma$. 
The electromagnetic boundary conditions at a conductive interface impose the continuity of the tangential electric field and a discontinuity of the tangential magnetic field proportional to the induced surface current density. Accordingly, the boundary conditions at the graphene sheet can be written as
\begin{subequations}\label{eq1}
\begin{align}
\hat{y} \times \left(\mathbf{H}_{\rm t} - \mathbf{H}_{\rm i} + \mathbf{H}_{\rm r}\right)\big|_{y=0}
&= \sigma \ast \mathbf{E}_{\rm t} = \mathbf{J}_{\rm s}, \label{eq1a}\\[0.15cm]
\hat{y} \times \left(\mathbf{E}_{\rm t} - \mathbf{E}_{\rm i} - \mathbf{E}_{\rm r}\right)\big|_{y=0}
&= 0,\label{eq1b}
\end{align}
\end{subequations}
where subscripts ${\rm i}$, ${\rm r}$, and ${\rm t}$ denote the incident, reflected, and transmitted fields, respectively, $\mathbf{J}_s$ is the surface current density, and ``$\ast$'' denotes convolution in time.
After straightforward algebraic manipulation, the above conditions can be recast into the scalar form
\begin{subequations}\label{eq2}
\begin{align}
\left(\frac{\eta_{2}+\eta_{1}}{2\,\eta_{1}}\right)E_{\rm t}
+ \frac{\eta_{2}}{2}\,\bigl\{\sigma\!\ast\!E_{\rm t}\bigr\}
= E_{\rm i}, \label{eq2a}\\[0.15cm]
E_{\rm t} - E_{\rm i} = E_{\rm r},\label{eq2b}
\end{align}
\end{subequations}
where $\eta_j=\sqrt{\mu_j/\varepsilon_j}$ is the wave impedance of medium $j = \{ 1, 2 \}$.

In a time-varying and dispersive material, the surface conductivity cannot be described by a time-invariant impulse response. Instead, the induced current at the observation time $t$ depends on the past history of the electric field, with a conductivity kernel that may vary explicitly with time. In this general case, the convolution term appearing in \eqref{eq2a} takes the form~\cite{Ptitcyn2023LPR, Mostafa2022PRApplied}
\[
\sigma \ast E_{\rm t} = \int_{-\infty}^{\infty} \sigma(t,t-t')\,E_{\rm t}(t')\,\mathrm{d}t',
\]
where $\sigma(t,t-t')$ encodes both material dispersion and explicit temporal modulation. Substituting this expression into \eqref{eq2a} yields the time-domain integral equation
\begin{equation}\label{eq3}
\left(\frac{\eta_{1}+\eta_{2}}{2\,\eta_{1}}\right)E_{\rm t}(t)
+ \frac{\eta_{2}}{2}\!\int_{-\infty}^{\infty}\!\!\sigma(t,t-t')\,E_{\rm t}(t')\,\mathrm{d}t'
= E_{\rm i}(t).
\end{equation}
It is worth noting that, in the absence of temporal modulation, the kernel reduces to $\sigma(t,t-t')=\sigma(t-t')$. In contrast, in a dispersionless but time-varying medium, it assumes the form $\sigma(t,t-t')=\sigma'(t)\delta(t-t')$. In the general dispersive and time-modulated case, however, the lack of time-translation invariance leads to frequency mixing. Taking the Fourier transform of \eqref{eq3} yields
\begin{equation}\label{eq4}
\left(\frac{\eta_{1}+\eta_{2}}{2\,\eta_{1}}\right)E_{\rm t}(\omega)
+ \frac{\eta_{2}}{2}\!\int_{-\infty}^{\infty}
\sigma(\omega-\omega',\omega')\,E_{\rm t}(\omega')\,\mathrm{d}\omega'
= E_{\rm i}(\omega).
\end{equation}

In \eqref{eq4}, a central feature of time-varying dispersive conductive interfaces is highlighted, namely that the electromagnetic response at a given frequency is coupled to neighboring spectral components through the operator $\sigma(\omega-\omega',\omega')$. To render the problem tractable, the frequency axis is discretized on a uniform grid $\omega_n=\omega_0+n\omega_{\rm s}$, where $\omega_0$ is the carrier frequency and $\omega_{\rm s}$ is a sampling frequency, chosen here equal to the modulation frequency. Under this discretization, the integral equation in \eqref{eq4} is reduced to the finite-dimensional system
\begin{equation}\label{eq5}
\left(\frac{\eta_{1}+\eta_{2}}{2\,\eta_{1}}\right)E_{\rm t}[n]
+ \frac{\eta_{2}}{2}\!\sum_{m=-K}^{K}\sigma[n-m]\,E_{\rm t}[m]
= E_{\rm i}[n].
\end{equation}
The coefficients $\sigma[n-m]$ quantify the amount of energy transferred from the harmonic component at frequency $\omega_m$ to the harmonic component at $\omega_n$. Collectively, \eqref{eq5} defines a system of $2K+1$ coupled equations, which can be compactly written for all sampled frequencies as
\begin{equation}\label{eq6}
\left(\frac{\eta_{1}+\eta_{2}}{2\,\eta_{1}}\mathbf{I}
+ \frac{\eta_{2}}{2}\mathbf{S}\right)\mathbf{E}^{(n)}_{\rm t}
= \mathbf{E}^{(n)}_{\rm i},
\end{equation}
where $\mathbf{I}$ is the identity matrix and $\mathbf{S}$ is the frequency-coupling matrix induced by the time-varying conductivity.
For a monochromatic excitation at $\omega_0$, only the central component of $\mathbf{E}^{(n)}_{\rm i}$ is non-zero, and the solution of \eqref{eq6} yields the full harmonic response of the graphene sheet. Once the transmitted field is obtained, the reflected field follows directly from \eqref{eq2b}. This operator-based formulation provides a rigorous and physically transparent description of the response of time-varying dispersive graphene sheets and serves as the foundation for the multilayer and resonant configurations analyzed in the remainder of this work.

\subsection{2.2 Taylor Expansion of the Drude Conductivity Model}\label{section2.2}
Let us now specialize the operator formulation developed in the previous subsection to graphene in the intraband regime, whose surface conductivity is accurately described by a Drude-type dispersion. Thus, the surface conductivity of graphene in the frequency domain can be written as~\cite{hanson2008}
\begin{equation}\label{eq7}
\sigma(\omega)
= \frac{e^2 k_B T}{\pi \hbar^2 (j\omega + 2\Gamma)}
\left(
\frac{\mu_{\rm c}}{k_B T}
+ 2 \ln \bigl( e^{-\mu_c/(k_B T)} + 1 \bigr)
\right),
\end{equation}
where $-e$ is the electron charge, $k_{\rm B}$ and $\hbar$ are the Boltzmann and reduced Planck constants, $\mu_{\rm c}$ is the chemical potential that can be controlled through an external electric bias field, $\Gamma$ is the energy-independent scattering rate, and $T$ is the temperature. In practice, $\mu_{\rm c}$ can be controlled electrostatically through an external gate bias. The applied bias alters the carrier density in graphene, which in turn shifts the Fermi level and therefore changes the chemical potential.
For compactness, we introduce the frequency-independent Drude prefactor
\begin{equation}\label{eq8}
A(\mu_{\rm c})
=
\frac{e^2}{\pi \hbar^2}
\left(
\mu_{\rm c} + 2 k_{\rm B} T \ln \bigl(1 + e^{-\mu_{\rm c}/(k_{\rm B} T)} \bigr)
\right),
\end{equation}
so that the conductivity assumes the simple form
\begin{equation}\label{eq9}
\sigma(\omega)=\frac{A(\mu_{\rm c})}{j\omega+2\Gamma}.
\end{equation}
Taking the inverse Fourier transform yields the causal time-domain response
\begin{equation}\label{eq10}
\sigma(t)=A(\mu_{\rm c})\,e^{-2\Gamma t}u(t),
\end{equation}
which explicitly separates the material memory kernel from the instantaneous dependence on the chemical potential. In previous works, a linear approximation is applied to \eqref{eq7} to remove the logarithmic factor and eventually simplify the analysis~\cite{Salary2018PRB, Sedeh2022TAP, Karamanos2024META}. However, this approximation is accurate only if $\mu_{\rm c} \gg k_{\rm B}T$, which corresponds approximately to $\mu_{\rm c}\approx 0.1\,$eV at room temperature $T=25^{\circ}{\textrm C}$. Accordingly, we first proceed, in this work, with a more general analysis, while the more common linearized approximation is summarized in \textit{Appendix A}.

Let us now assume that the conductivity can be controlled through a temporal modulation of the chemical potential,
\begin{equation}\label{eq11}
\mu_{\rm c}(t)=\mu_{{\rm c},0}+\Delta\mu\cos\theta,
\end{equation}
where $\theta = \omega_{\mathrm{mod}} t$ and $\omega_{\mathrm{mod}}$ is the modulation frequency, $\mu_{{\rm c},0}$ and $\Delta\mu$ denote the chemical potential due to the constant and time-modulated bias components, respectively. In the case of $\mu_{{\rm c},0} > 0$, the modulation depth $M$ can be introduced to connect the two chemical potential variables:
\begin{equation}\label{eq12}
\Delta\mu = \mu_{{\rm c},0} M.
\end{equation}

A key observation is that the Drude prefactor $A(\mu)$ is analytic in a neighborhood of the chosen bias point $\mu_{{\rm c},0}$. Consequently, its temporal modulation can be expressed locally through a Taylor expansion around $\mu_{{\rm c},0}$ as,
\begin{equation}\label{eq13}
A(\mu_{{\rm c},0}+\Delta\mu\cos\theta)
=
\sum_{p=0}^{\infty}
\frac{A^{(p)}(\mu_{{\rm c},0})}{p!}
\bigl(\Delta\mu\cos\theta\bigr)^p,
\end{equation}
where $A^{(p)}(\mu)$ denotes the $p$-th derivative with respect to $\mu$.
At this point, it is important to emphasize that the expansion in \eqref{eq13} is local in the complex $\mu$-plane. In particular, standard complex-analysis theory implies that an analytic function admits a convergent power-series expansion inside a disk centered at the expansion point, with radius determined by the distance to the nearest singularity. In the present case, the logarithmic term in \eqref{eq8} introduces complex singularities through the condition 
\begin{equation}\label{eq14}
1+e^{-\mu/(k_{\rm B} T)}=0,
\end{equation}
which yields the sequence
\begin{equation}\label{eq15}
\mu=(2\ell+1)i\pi k_{\rm B} T,\qquad \ell\in\mathbb{Z}.
\end{equation}
Therefore, for a real bias point $\mu_{{\rm c},0}$, the nearest singularities are located at $\mu=\pm i\pi k_{\rm B} T$, and the corresponding radius of convergence is 
\begin{equation}\label{eq16}
    R(\mu_{{\rm c},0})=\sqrt{\mu_{{\rm c,}0}^{\,2}+(\pi k_{\rm B} T)^2}.
\end{equation}
Accordingly, the Taylor-expanded representation is expected to remain valid provided that the modulation excursion satisfies $\Delta\mu<R(\mu_{{\rm c},0})$. A brief derivation of this result is provided in \emph{Appendix B}.

The derivatives of $A(\mu)$ exhibit a well-defined structure. By introducing the normalized variable $x=\mu/(k_{\rm B} T)$ and the Fermi-Dirac function $f(x)=1/(1+e^{x})$, one finds that each successive derivative introduces an additional factor of $1/(k_\textbf{B} T)$ and polynomial dependence on $f$. In particular, for all integers $n\geq2$, the derivatives can be written in the compact form
\begin{equation}\label{eq17}
A^{(n)}(\mu)
=
\frac{2e^2}{\pi\hbar^2}
\frac{1}{(k_{\rm B} T)^{n-1}}
f(1-f)\,P_{n-2}(f),
\end{equation}
where $P_{n-2}(f)$ is a polynomial generated recursively. Since $f'=-f(1-f)$, the polynomial sequence is initialized by $P_0(f)=1$, which corresponds to the second derivative $A''(\mu)$, and satisfies the recursion
\begin{equation}\label{eq18}
P_{m+1}(f)
=
-\frac{\mathrm{d}}{\mathrm{d}f}\!\left[f(1-f)\,P_m(f)\right]
=
-(1-2f)P_m(f)-f(1-f)\frac{\mathrm{d}P_m}{\mathrm{d}f},
\qquad m\geq0.
\end{equation}
The first few members of the sequence are therefore
\begin{equation}\label{eq19}
P_0(f)=1,\qquad
P_1(f)=2f-1,\qquad
P_2(f)=1-6f+6f^2,
\end{equation}
which illustrate the polynomial structure generated by successive differentiation.

Each Taylor term of order $p$ contains the factor $\cos^p\theta$, which can be decomposed into a finite sum of harmonics as
\begin{equation}\label{eq20}
\cos^p\theta
=
\frac{1}{2^p}
\sum_{m=0}^{p}
\binom{p}{m}
e^{j(p-2m)\theta},
\end{equation}
where $m$ is the binomial index. By conjugate symmetry, this expansion yields a purely real Fourier series. Collecting all terms oscillating at the same multiple of $\omega_{\mathrm{m}}$, the Drude prefactor can therefore be written as a finite harmonic expansion. Truncating the Taylor series at order $p=P$ gives
\begin{equation}\label{eq21}
A(t)\equiv A(\mu_{\rm c}(t))
\simeq
\sum_{n=0}^{P} A_n \cos(n\theta),
\end{equation}
where $n$ denotes the harmonic order, and the corresponding coefficients are
\begin{equation}\label{eq22}
A_n
=
\sum_{q=0}^{\lfloor (P-n)/2 \rfloor}
\frac{2-\delta_{n0}}{2^{\,n+2q}}
\binom{n+2q}{q}
\frac{A^{(n+2q)}(\mu_{{\rm c},0})}{(n+2q)!}
(\Delta\mu)^{\,n+2q},
\qquad n=0,1,\dots,P.
\end{equation}
Here, $q$ is a non-negative integer indexing the Taylor orders $p=n+2q$ that contribute to the $n$-th modulation harmonic, and $\delta_{n0}$ denotes the Kronecker delta. This expression shows explicitly that odd Taylor orders contribute only to odd modulation harmonics, whereas even Taylor orders contribute only to even harmonics. Consequently, truncation at order $P$ produces a finite Floquet bandwidth extending up to $\pm P\omega_{\mathrm{mod}}$.
 
The Fourier transform of the truncated expansion of \eqref{eq22} can be written as a weighted impulse train, 
\begin{equation}\label{eq23}
A(\omega)
=
2\pi\sum_{n=-P}^{P}
\bar{A}_n\,\delta(\omega-n\omega_{\mathrm{mod}}),
\end{equation}
with $\bar{A}_0=A_0$ and $\bar{A}_{\pm n}=A_n/2$ for $n\geq1$. Therefore, by substituting \eqref{eq22} into \eqref{eq10}, the resulting frequency-domain conductivity is 
\begin{equation}\label{eq24}
\sigma(\omega)
=
2\pi\sum_{n=-P}^{P}
\frac{\bar{A}_n}{j(\omega-n\omega_{\mathrm{mod}})+2\Gamma}.
\end{equation}

The Taylor-expanded formulation provides a general and physically transparent description of temporal conductivity modulation. Importantly, it remains applicable even when the modulation range spans both positive and negative values of the chemical potential, as may occur under strong bias reversal. In particular, because the Drude prefactor $A(\mu_{\rm c})$ is an even function of the chemical potential, a modulation centered about $\mu_{{\rm c},0}=0$ gives rise to purely even harmonic generation. This feature cannot be captured by linearized models~\cite{Salary2018PRB, Sedeh2022TAP} (see \textit{Appendix A}). More generally, when the modulation crosses zero but is not centered at $\mu_{{\rm c},0}=0$, odd harmonics may still be present. This generality is essential for correctly predicting the harmonic content of strongly modulated graphene. 

\subsection{2.3 Transfer-Matrix Model for Multilayer Graphene Cavities}\label{subsection2.3}

\begin{figure*}
    \centering
    \includegraphics[width=0.85\linewidth]{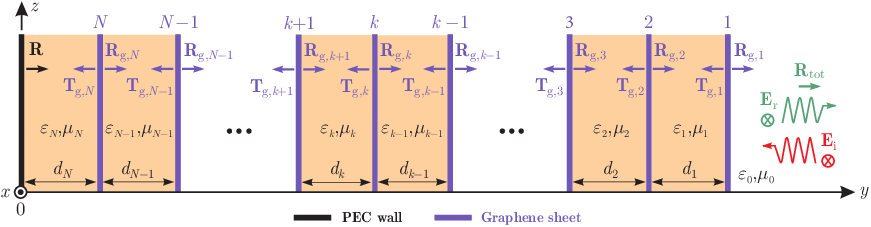}
    \caption{The time-modulated graphene resonator array composed of $N$ graphene sheets and terminated by a PEC wall.}
    \label{fig2}
\end{figure*}

Let us now extend the single-sheet response derived in the previous sections to multilayer configurations composed of multiple time-modulated graphene sheets separated by dielectric spacers. The structure under consideration consists of $N$ graphene sheets placed successively along one direction and terminated on the other side by a PEC surface, forming a Fabry--Perot-like resonator, as depicted in Figure~\ref{fig2}. Each graphene sheet $k$ is separated from its neighboring sheet $k-1$ by a dielectric layer of thickness $d_k$ and material parameters $(\varepsilon_k,\mu_k)$.

At any reference plane between adjacent layers, the field is decomposed into right-propagating and left-propagating Floquet amplitude vectors, denoted by $\mathbf{E}^+$ and $\mathbf{E}^-$. Let $\mathbf{E}_{k-1}^{\pm}$ and $\mathbf{E}_{k}^{\pm}$ denote the amplitudes immediately to the left and right of the $k$-th graphene sheet, respectively. In the general case where the media on the two sides of the sheet are different, the Floquet scattering operators are direction-dependent. Thus, the scattering relations across the $k$-th graphene sheet are written as
\begin{subequations}\label{eq25}
\begin{align}
\mathbf{E}_{k-1}^{-}
&=
\mathbf{R}_{{\rm g},k}^{\mathrm{L}}\,\mathbf{E}_{k-1}^{+}
+
\mathbf{T}_{{\rm g},k}^{\mathrm{R}\rightarrow \mathrm{L}}\,\mathbf{E}_{k}^{-},
\label{eq25a}\\
\mathbf{E}_{k}^{+}
&=
\mathbf{T}_{{\rm g},k}^{\mathrm{L}\rightarrow \mathrm{R}}\,\mathbf{E}_{k-1}^{+}
+
\mathbf{R}_{{\rm g},k}^{\mathrm{R}}\,\mathbf{E}_{k}^{-},
\label{eq25b}
\end{align}
\end{subequations}
where $\mathbf{R}_{{\rm g},k}^{\mathrm{L}}$ and $\mathbf{R}_{{\rm g},k}^{\mathrm{R}}$ denote the reflection operators of each graphene sheet for incidence from the left and right sides of the sheet, respectively, while $\mathbf{T}_{{\rm g},k}^{\mathrm{L}\rightarrow \mathrm{R}}$ and $\mathbf{T}_{{\rm g},k}^{\mathrm{R}\rightarrow \mathrm{L}}$ denote the corresponding transmission operators. These operators are obtained from the single-sheet operator equation in \eqref{eq6}.

Solving \eqref{eq25} for the field amplitudes on the right side of the sheets gives
\begin{equation}\label{eq26}
\begin{pmatrix}
\mathbf{E}_k^+ \\[2mm]
\mathbf{E}_k^-
\end{pmatrix}
=
\boldsymbol{\mathcal{L}}_k
\begin{pmatrix}
\mathbf{E}_{k-1}^+ \\[2mm]
\mathbf{E}_{k-1}^-
\end{pmatrix},
\end{equation}
with the interface transfer matrix
\begin{equation}\label{eq27}
\boldsymbol{\mathcal{L}}_k
=
\begin{pmatrix}
\mathbf{T}_{{\rm g},k}^{\mathrm{L}\rightarrow \mathrm{R}}
-
\mathbf{R}_{{\rm g},k}^{\mathrm{R}}
\big(\mathbf{T}_{{\rm g},k}^{\mathrm{R}\rightarrow \mathrm{L}}\big)^{-1}
\mathbf{R}_{{\rm g},k}^{\mathrm{L}}
&
\mathbf{R}_{{\rm g},k}^{\mathrm{R}}
\big(\mathbf{T}_{{\rm g},k}^{\mathrm{R}\rightarrow \mathrm{L}}\big)^{-1}
\\[2mm]
-
\big(\mathbf{T}_{{\rm g},k}^{\mathrm{R}\rightarrow \mathrm{L}}\big)^{-1}
\mathbf{R}_{{\rm g},k}^{\mathrm{L}}
&
\big(\mathbf{T}_{{\rm g},k}^{\mathrm{R}\rightarrow \mathrm{L}}\big)^{-1}
\end{pmatrix}.
\end{equation}
In the symmetric case where the media on the two sides of the sheet are identical, one has
$\mathbf{R}_{{\rm g},k}^{\mathrm{L}}=\mathbf{R}_{{\rm g},k}^{\mathrm{R}}\equiv \mathbf{R}_{{\rm g},k}$ and
$\mathbf{T}_{{\rm g},k}^{\mathrm{L}\rightarrow \mathrm{R}}=\mathbf{T}_{{\rm g},k}^{\mathrm{R}\rightarrow \mathrm{L}}\equiv \mathbf{T}_{{\rm g},k}$,
so that \eqref{eq27} reduces to
\begin{equation}\label{eq28}
\boldsymbol{\mathcal{L}}_k
=
\begin{pmatrix}
\mathbf{T}_{{\rm g},k}-\mathbf{R}_{{\rm g},k}\mathbf{T}_{{\rm g},k}^{-1}\mathbf{R}_{{\rm g},k}
&
\mathbf{R}_{{\rm g},k}\mathbf{T}_{{\rm g},k}^{-1}
\\[2mm]
-\mathbf{T}_{{\rm g},k}^{-1}\mathbf{R}_{{\rm g},k}
&
\mathbf{T}_{{\rm g},k}^{-1}
\end{pmatrix}.
\end{equation}
In the cavity configurations considered here, this symmetric reduction applies to the interior graphene sheets when the cavity medium is uniform, whereas the exterior-facing sheet generally requires the full directional formulation if the exterior and cavity media differ.

Propagation through the $k$-th dielectric spacer, situated between the graphene sheets $k$ and $k-1$ (Fig.~\ref{fig2}), is described by the diagonal matrix
\begin{equation}\label{eq29}
\boldsymbol{\mathcal{P}}_k
=
\begin{pmatrix}
\boldsymbol{\mathcal{Q}}_k & 0 \\
0 & \boldsymbol{\mathcal{Q}}_k^{-1}
\end{pmatrix},
\qquad
\boldsymbol{\mathcal{Q}}_k
=
\mathrm{diag}\!\left\{e^{-j k_n^{(k)} d_k}\right\},
\end{equation}
where $d_k$ is the thickness of the $k$-th spacer and
$k_n^{(k)}=\omega_n\sqrt{\mu_k\varepsilon_k}$ is the wavenumber of the $n$-th Floquet harmonic in that spacer medium.

Combining interface and propagation effects, from \eqref{eq27},\eqref{eq28} and \eqref{eq29}, respectively, the transfer matrix of the $k$-th unit cell is calculated as
\begin{equation}\label{eq30}
\mathbf{M}_k
=
\boldsymbol{\mathcal{P}}_k\,\boldsymbol{\mathcal{L}}_k.
\end{equation}
Therefore, the total transfer matrix of the structure follows as
\begin{equation}\label{eq31}
\mathbf{M}_{\mathrm{tot}}
=
\mathbf{M}_N \mathbf{M}_{N-1} \cdots \mathbf{M}_1
=
\begin{pmatrix}
\mathbf{A} & \mathbf{B} \\
\mathbf{C} & \mathbf{D}
\end{pmatrix},
\end{equation}
which relates the input-plane amplitudes to those at the PEC termination plane according to
\begin{equation}\label{eq32}
\begin{pmatrix}
\mathbf{E}_N^+ \\
\mathbf{E}_N^-
\end{pmatrix}
=
\mathbf{M}_{\mathrm{tot}}
\begin{pmatrix}
\mathbf{E}_0^+ \\
\mathbf{E}_0^-
\end{pmatrix}.
\end{equation}
The PEC boundary condition requires the total tangential electric field to vanish at the terminating wall, namely
\begin{equation}\label{eq33}
\mathbf{E}_N^+ + \mathbf{E}_N^- = 0.
\end{equation}
Setting $\mathbf{E}_0^+ \equiv \mathbf{E}_i$ and $\mathbf{E}_0^- \equiv \mathbf{E}_r$ and substituting the block form of $\mathbf{M}_{\mathrm{tot}}$ into this condition yields
\begin{equation}\label{eq34}
(\mathbf{A}+\mathbf{C})\mathbf{E}_i + (\mathbf{B}+\mathbf{D})\mathbf{E}_r = 0,
\end{equation}
and, finally, the overall reflection operator for the Fabry--Perot-like cavity of Fig.~\ref{fig2} becomes
\begin{equation}\label{eq35}
\mathbf{R}_{\mathrm{tot}}
=
-(\mathbf{B}+\mathbf{D})^{-1}(\mathbf{A}+\mathbf{C}),
\qquad
\mathbf{E}_r=\mathbf{R}_{\mathrm{tot}}\mathbf{E}_i.
\end{equation}
This formulation provides a compact and systematic framework for computing the reflected harmonic spectrum of multilayer time-modulated graphene structures, while remaining applicable to both uniform and nonuniform dielectric stacks.

The analysis of such multilayer systems can be carried out using two equivalent but complementary approaches. In addition to the transfer matrix method presented above, a recursive multiple-scattering formulation can also be employed to interpret the response in terms of successive internal reflections within the structure. Moreover, both approaches rely on the Floquet reflection and transmission operators of each graphene sheet. For completeness, this alternative derivation is provided in \textit{Appendix C}, while the transfer matrix formulation is adopted for the rest of the analysis due to its compactness and computational efficiency.

\subsection{2.4 Full-Wave Validation via FDTD}

To validate the proposed semi-analytic framework developed in the previous sections, we benchmark the transfer-matrix results against an independent, in-house finite-difference time-domain (FDTD) solver. In this FDTD implementation, graphene is modeled as an equivalent surface current density $\mathbf{J}_{\rm g} = \sigma_\mathrm{\rm g}\, \mathbf{E}$, with temporal dispersion incorporated through an auxiliary differential equation (ADE) scheme~\cite{Karamanos2024META}, resulting in the discrete update equation of the surface current
\begin{equation}\label{eq36}
J_{{\rm g},k}\big|^{\,n+\frac{1}{2}}
= \frac{1-\Gamma \Delta t}{1+\Gamma \Delta t}\,J_{{\rm g},k}\big|^{\,n-\frac{1}{2}}
+ \frac{A_{\mu_{c,0}}\big|^{\,n}}{1+\Gamma \Delta t}\,\Delta t 
\,E_z\big|^{\,n}_k,
\end{equation}
where the frequency-independent term $A_{\mu_{{\rm c},0}}$ is updated using \eqref{eq8} after applying the temporal modulation in chemical potential:
\begin{equation}\label{eq37}
\mu_{\rm c}\big|^{\,n} = \mu_{{\rm c},0} + \Delta \mu \, \cos (\omega_{\rm mod}\, n\, \Delta t).
\end{equation}
Then, \eqref{eq36} is incorporated into the FDTD expressions in terms of Ampere's law, namely
\begin{equation}\label{eq38}
\begin{aligned}
E_z\big|^{\,n+1}_i
&= E_z\big|^{\,n}_i
+ \frac{\Delta t}{\varepsilon_i\,\Delta x}
\Big[ H_y\big|^{\,n+\frac{1}{2}}_{i+\frac{1}{2}} 
- H_y\big|^{\,n+\frac{1}{2}}_{i-\frac{1}{2}} \Big] - \frac{\Delta t}{\varepsilon_i\,\Delta x} J_{{\rm g},i}\big|^{\,n+\frac{1}{2}}.
\end{aligned}
\end{equation}
This approach enables direct time-domain simulation of dispersive, time-modulated conductive interfaces and serves as a stringent reference for validation.

\begin{figure*}
  \centering
\begin{subfigure}[t]{0.48\textwidth}
    \centering
    \includegraphics[width=\linewidth]{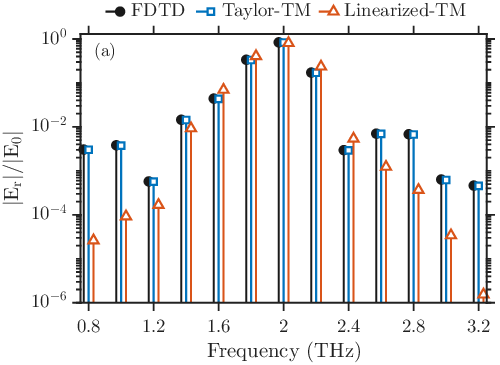}
    \caption{$\mu_{c,0}=0.1\,$eV, $\Delta\mu=0.12\,$eV}
    \label{fig3a}
  \end{subfigure}\hfill
  \begin{subfigure}[t]{0.48\textwidth}
    \centering
    \includegraphics[width=\linewidth]{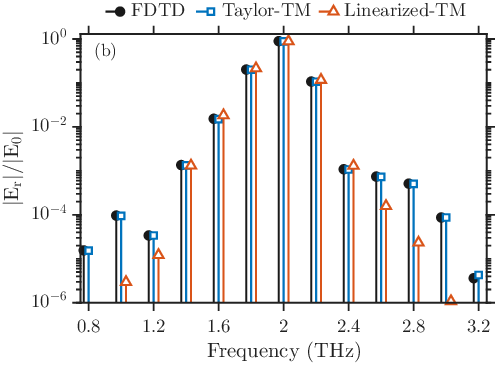}
    \caption{$\mu_{c,0}=0.1\,$eV, $\Delta\mu=0.06\,$eV}
    \label{fig3b}
  \end{subfigure}\hfill
  \begin{subfigure}[t]{0.48\textwidth}
    \centering
    \includegraphics[width=\linewidth]{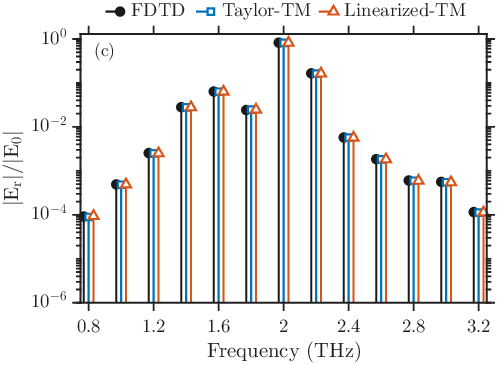}
    \caption{$\mu_{c,0}=0.5\,$eV, $\Delta\mu=0.3\,$eV}
    \label{fig3c}
  \end{subfigure}

\caption{Validation of the transfer matrix method with the Taylor-expanded conductivity model (Taylor--TM) and the transfer matrix method with the linearized conductivity approximation (Linearized--TM) versus simulations via the modified FDTD algorithm.
The three cases correspond to $\mu_{\rm c}(t)\in[-0.02,0.22]\,$eV, $\mu_{\rm c}(t)\in[0.04,0.16]\,$eV, and $\mu_{\rm c}(t)\in[0.2,0.8]\,$eV, respectively.}
  \label{fig3}
  \vspace{-0.3cm}
\end{figure*}

All validation results correspond to a stratified structure composed of $N=3$ time-varying graphene sheets separated by equal dielectric spacings ($d=0.12441\,$mm) and terminated by a PEC wall. A monochromatic plane wave at $f_0 = 2\,$THz illuminates the structure under normal incidence. The modulation frequency for the graphene sheets is chosen as $\omega_{\rm mod}=\omega_0/10$, while the sampling frequency is set equal to the modulation frequency, i.e., $\omega_{s} = \omega_{\rm mod}$. The reflected harmonic amplitudes are extracted on a logarithmic scale to enable comparison over several orders of magnitude. The three cases reported for validation in Fig.~\ref{fig3} were selected to span mildly sign-crossing, moderately nonlinear, and high-bias regimes. In particular, the first case provides a stringent nonlinear test while remaining consistent with the local convergence condition of the Taylor-expanded conductivity model discussed in \textit{Section}~2.2 and \textit{Appendix~B}.

Figure~\ref{fig3} presents the reflected harmonic spectra obtained using full-wave FDTD simulations and the proposed transfer matrix framework under three representative modulation conditions. In Fig.~\ref{fig3}(a), the chemical potential is weakly sign-crossing, so that the conductivity response already exhibits a distinctly nonlinear character. In this regime, the Taylor-expanded transfer-matrix formulation remains in very good agreement with FDTD, whereas the linearized approximation shows noticeable deviations, especially for the generated higher-order harmonics. This indicates that higher-order temporal mixing is already relevant and cannot be captured accurately by a first-order approximation model for graphene's conductivity. 

Next, Fig.~\ref{fig3}(b) corresponds to a more moderate modulation excursion about the same bias point. In this case, the Taylor-based transfer-matrix results continue to agree very well with FDTD across the spectrum, while the linearized approximation becomes substantially more accurate, especially for the dominant harmonics. The remaining discrepancies are mainly confined to weaker higher-order sidebands, reflecting the fact that the first-order approximation retains only nearest-neighbor Floquet coupling.
Finally, Fig.~\ref{fig3}(c) shows the high-bias case $\mu_{{\rm c},0}=0.5\,$eV. Here, the Drude prefactor in \eqref{eq8} is effectively linear versus the chemical potential (see \textit{Appendix A}), and the conductivity dynamics are dominated by nearest-neighbor harmonic coupling. As a result, all three approaches, FDTD, the Taylor-based transfer-matrix, and the linearized transfer-matrix approximation, are in near-perfect agreement throughout the spectrum. In this regime, the sparse tridiagonal coupling structure assumed by the approximation becomes fully valid, confirming both its accuracy and its computational efficiency for high-bias operation.

\section{3. Optimization-Driven Harmonic Engineering in Time-Modulated Graphene Cavities}

Having established and validated the proposed semi-analytic framework, we now employ it for optimization-driven harmonic engineering in multilayer time-modulated graphene cavities. The structure under study consists of $N$ time-modulated graphene sheets and terminated by a PEC wall, as depicted in Fig.~\ref{fig2}. The cavity region between the sheets is filled with dielectric spacers of relative permittivity $\varepsilon_r$, while the exterior region is assumed to be air. Note that the first graphene sheet forms an asymmetric interface, due to the difference of the exterior region and the cavity medium. Accordingly, the scattering operators are evaluated using the directional expressions of \textit{Section}~2.3, whereas the symmetric cavity-to-cavity form describes the remaining interior sheets. As for the constant graphene parameters, scattering rate is selected $\Gamma=0.11\,$meV at the room's temperature, namely $T=300\,$K. Addtionally, all scenarios share the same excitation frequency at $f_{0}=2\,$THz with a modulation frequency $f_{\rm mod}=0.2\,$THz. Finally, the optimization is carried out using the particle swarm optimization (PSO) algorithm \cite{GlobalOptimizationToolbox}. The design variables correspond either to a single common cavity spacing or to a set of independent gaps $\mathbf{d}=\{d_k\}$, depending on the considered scenario. In all cases, the optimization is performed under physically realistic bounds on the cavity spacings, which are specified in the corresponding subsections. Since PSO is a stochastic global optimization method, convergence is verified through repeated runs to ensure the robustness of the obtained solutions. Although other optimization strategies could be employed, PSO is selected here due to its simplicity and effectiveness for the present non-convex design problem.

\subsection{3.1 First Generated Harmonic Enhancement in the High-Bias Regime}
In this first example, we operate in the high-bias regime, where the constant chemical potential of graphene is set to $\mu_{{\rm c},0}=0.5\,$eV, with the modulated one $\Delta\mu=0.3$ eV (modulation depth $M=0.6$). Using this bias conditions, the Drude prefactor is effectively linear over the modulation excursion coinciding to the high-bias validation case of Fig.~\ref{fig3c}. As a result, the linearized conductivity model summarized in \textit{Appendix A} provides an accurate and computationally efficient description of the harmonic coupling. Hence, the optimization is carried out using the approximate linearized transfer-matrix formulation. Concerning the geometric parameters, the structure consists of $N=10$ graphene sheets separated by silicon dioxide (SiO$_2$) dielectric spacers with $\varepsilon_{\rm r}=3.8$.

The design objective is to selectively enhance a chosen Floquet harmonic by tuning the common cavity spacing $d$ for all unit cells. Denoting by $\mathbf{e}_n$ the canonical basis vector selecting the $n$-th harmonic, the reflected spectrum can be written as $\mathbf{E}_{\rm r}(d)=\mathbf{R}_{\mathrm{tot}}(d)\mathbf{e}_{n,0}$, where $n_0$ corresponds to the central harmonic. To maximize the amplitude of a target harmonic $n_t$, we define the optimization objective function to be minimized as
\begin{align}\label{eq39}
    f(d) = -\lvert \mathbf{e}_{n,t}^{\mathsf T}\,\mathbf{E}_{\,{\rm r}}(d) \rvert
    = -\lvert \mathbf{e}_{n,t}^{\mathsf T}\,\mathbf{R}_{\mathrm{tot}}(d)\,\mathbf{e}_{n,0} \rvert.
\end{align}

\begin{figure*}
  \centering
  \begin{subfigure}[t]{0.48\textwidth}
    \centering
    \includegraphics[width=\linewidth]{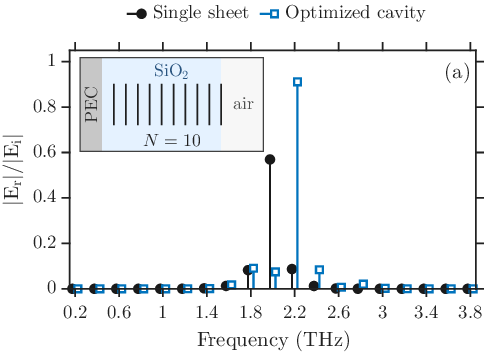}
    \caption{Optimized enhancement of the upper first generated harmonic $f_{+1}=f_0+f_{\rm mod}=2.2\,$THz.}
    \label{fig4a}
  \end{subfigure}\hfill
  \begin{subfigure}[t]{0.48\textwidth}
    \centering
    \includegraphics[width=\linewidth]{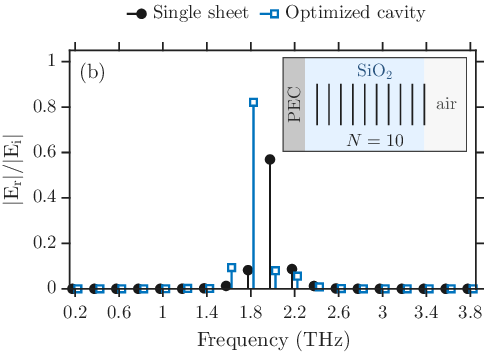}
    \caption{Optimized enhancement of the lower first generated harmonic $f_{-1}=f_0-f_{\rm mod}=1.8\,$THz.}
    \label{fig4b}
  \end{subfigure}

  \caption{Selective enhancement of first generated harmonics, or $f_{\pm 1}=f_0 \pm f_{\rm mod}$, in a PEC-terminated stack of time-modulated graphene sheets using cavity optimization.
The single-sheet response is compared with the optimized multi-reflection one obtained via the approximate semi-analytic transfer-matrix model.}
  \label{fig4}
\end{figure*}

In this scenario, the graphene sheets are assumed equally spaced with the optimization performed over one modulation wavelength in the cavity medium, \(d\in[0,\lambda_{\rm mod}^{(\mathrm{SiO_2})}]\), where \(\lambda_{\rm mod}^{(\mathrm{SiO_2})}=c_0/(f_{\rm mod}\sqrt{\varepsilon_r})\). We first target the upper first generated harmonic, $f_{+1}=f_0+f_{\rm mod}=2.2\,$THz. The optimization yields an optimal cavity gap of $d_{\mathrm{opt}}= 1.74\lambda_0 = 0.2600788\,$mm. For this configuration, the normalized reflected amplitude at the $+1$ harmonic increases from the single-sheet reference value $|E_{\rm r}/E_{\rm i}|=0.087$ to $|E_{\rm r}/E_{\rm i}|=0.911$. This corresponds to an enhancement factor of $10.48$, or an amplitude gain of $20.41\,$dB. The corresponding spectrum is shown in Fig.~\ref{fig4a}, where the strong and selective amplification of the target harmonic is clearly observed.
We then repeat the same procedure for the lower first-order sideband at $f_{-1}=f_0-f_{\rm mod}=1.8\,$THz. In this case, the PSO algorithm yields an optimal spacing $d_{\mathrm{opt}}= 3.27\lambda_0 = 0.4904987\,$mm. The reflected amplitude at the $-1$ harmonic increases from $|E_{\rm r}/E_{\rm i}|=0.082$ in the single-sheet configuration to $|E_{\rm r}/E_{\rm i}|=0.821$ after optimization. This corresponds to an enhancement factor of $9.98$, or an amplitude gain of $19.99\,$dB. The resulting spectrum is shown in Fig.~\ref{fig4b}.

In both cases, the enhancement mechanism can be traced back to a phase-matching condition imposed by the optimized cavity spacing, where the round-trip phase accumulated by the target harmonic is aligned such that multiple reflections from the graphene stack interfere constructively. Using the Taylor-based transfer-matrix formulation (see \textit{Section}~2.2) would yield indistinguishable spectra for this bias level, because the observed enhancement is governed by first-order harmonic coupling and higher-order conductivity nonlinearities are negligible in this regime, as demonstrated in \textit{Section}$~2.4$.

\begin{figure}[t]
  \centering
  \includegraphics[width=0.5\linewidth]{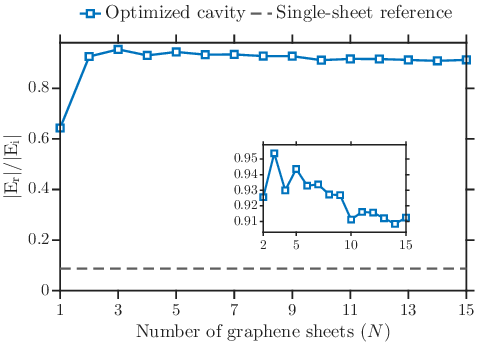}
  \caption{Parametric study of the optimized upper first generated harmonic, or $f_{+1}=f_0+f_{\rm mod}=2.2\,$THz, as a function of the number of graphene sheets $N$. The solid curve denotes the optimized PEC-terminated cavity response, while the dashed line indicates the isolated single-sheet reference. The inset zooms into the range $N\geq 2$ and shows the rapid onset of saturation once a short multilayer cavity is formed. The remaining setup parameters are the same as in Fig.~\ref{fig4a}.}
  \label{fig5}
\end{figure}

To further clarify the role of the unit-cell number in first-generated harmonic augmentation, we next repeat the optimization for the upper sideband at $f_{+1}=f_0+f_{\rm mod}=2.2\,$THz while varying the number of graphene sheets. In this parametric study, the number of graphene sheets is swept from $N=1$ to $N=15$, and for each value of $N$, the common cavity spacing $d$ is re-optimized over the interval $[0,\lambda_{\rm mod}^{(\mathrm{SiO_2})}]$ using the same objective function as in \eqref{eq39}. The resulting amplitudes from the PSO algorithm are summarized in Fig.~\ref{fig5}, where they are compared against the isolated single-sheet reference.
For $N=1$, the optimized PEC-terminated cavity already yields a reflected amplitude of $|E_{\rm r}/E_{\rm i}|=0.643$ at the target sideband, corresponding to an enhancement factor of $7.39$, or an amplitude gain of $17.38\,$dB relative to the single-sheet response. Increasing the cavity order to $N=2$ and $N=3$ raises the optimized amplitude further to $|E_{\rm r}/E_{\rm i}|=0.926$ and $|E_{\rm r}/E_{\rm i}|=0.954$, respectively. Beyond this point, however, the improvement becomes weak. For $N=4$ to $N=15$, the optimized response remains confined to the narrow interval $|E_{\rm r}/E_{\rm i}| \approx 0.906$-$0.944$, which corresponds to enhancement factors between $10.42$ and $10.85$, or amplitude gains between $20.35$ and $20.71\,$dB. The inset of Fig.~\ref{fig5} highlights these small variations about the saturation plateau.
This behavior indicates that, within the design depicted in Fig.~\ref{fig2}, almost the maximum possible enhancement of the first harmonics is captured already when the structure evolves from a single modulated sheet to a short cavity of a few unit cells or graphene layers. Additional graphene sheets still allow the optimizer to recover phase-matched solutions, but they do not lead to substantial further growth of the targeted frequency. In other words, under the present constant bias chemical potential, $\mu_{{\rm c},0}$, and common spacing constraints, $d$, the number of graphene sheets acts as an enhancement parameter mainly for small values of $N$, whereas for larger cavities, the achievable enhancement becomes only weakly dependent on the total number of layers. This trend is consistent with the first-order coupling picture of the linearized model, in which the dominant mechanism is the constructive build-up of nearest-neighbor Floquet conversion assisted by internal cavity reflections.

\subsection{3.2 Third Generated Frequency Enhancement Beyond the Linear Regime}
Let us now consider a second design paradigm that operates beyond the regime where the linearized conductivity model can be relied upon for harmonic-generation engineering. In this example, the goal is to selectively enhance the upper third generated harmonic, or $f_{+3}=f_0+3f_{\rm mod}=2.6\,$THz. In this scenario a lower constant chemical potential is used, namely $\mu_{{\rm c},0}=0.2\,$eV, retaining the modulation depth $M=0.6$ leading in a modulation excursion $\Delta\mu=0.12\,$eV. Thus, the chemical potential varies over the interval $\mu_{\rm c}(t)\in[0.08,0.32]\,$eV, which is outside the high-bias regime, constituting the first-order linearized approximation, described in \textit{Appendix A}, not suitable for an accurate representation. At the same time, it remains within the local convergence disk of the Taylor-expanded Drude prefactor, since for $\mu_{{\rm c},0}=0.2\,$eV and $T=25^\circ$C one obtains $R(\mu_{{\rm c},0})=\sqrt{\mu_{{\rm c},0}^2+(\pi k_{\rm B} T)^2}\approx 0.216\,$eV $>\Delta\mu$. As a result, the design and optimization are carried out using the Taylor-expanded transfer matrix formulation (see \textit{Section}~2.3). Finally, this structure consists of $N=15$ time-modulated graphene sheets separated by polytetrafluoroethylene (PTFE) spacers with $\varepsilon_r=2$.

As in the previous section, the reflected spectrum is written as $\mathbf{E}_{\rm r}(d)=\mathbf{R}_{\rm tot}(d)\,\mathbf{e}_{n,0}$, where $d=\{d_1,d_2,\dots,d_{15}\}$ denotes the set of cavity gaps, which are now allowed to vary independently. To enforce not only a large amplitude at the target harmonic but also suppression of spurious sidebands, we adopt a composite objective function of the form \cite{christiansen2021inverse}
\begin{align}\label{eq40}
   J(d) = \left[\sum\nolimits_{n, n\neq t} \sqrt{w\,|E_{\rm r}^{(n)}(\mathbf{d})|}  \right]^{2}
   + \Big(|E_{\rm r}^{(n,t)}(\mathbf{d})|\Big)^{-1},
\end{align}
where $n_t=n_0+3$ denotes the target harmonic index, $E_{\rm r}^{(n)}$ is the $n$-th component of $\mathbf{E}_{\rm r}$, and $w$ is a weighting parameter that controls the trade-off between peak amplitude and the suppression of all other generated harmonics. In the results that follow, we set $w=1$, which was found to provide a balanced compromise between strong enhancement and effective suppression of non-target harmonics.

As a reference example, we first consider a baseline optimization in which the cavity gaps $\{d_k\}_{k=1}^{15}$ are optimized independently under the bounds $0.2\lambda_0 \le d_k \le \lambda_0$, while the single-objective function in \eqref{eq39} is used to maximize only the target harmonic. Under these constraints, the reflected amplitude at the third generated harmonic reaches $|E_{\rm r}|=0.625$, corresponding to an enhancement of $62.08\,$dB relative to the single-sheet response. However, the resulting spectrum remains relatively broad, with significant energy still distributed among neighboring generated harmonics, most notably the upper first and second ones. This confirms that the maximization of the target harmonic alone does not guarantee a spectrally isolated response, even when the cavity geometry is optimized freely within realistic bounds.

\begin{figure*}[t!]
  \centering
  \begin{subfigure}[t]{0.48\textwidth}
    \centering
    \includegraphics[width=\linewidth]{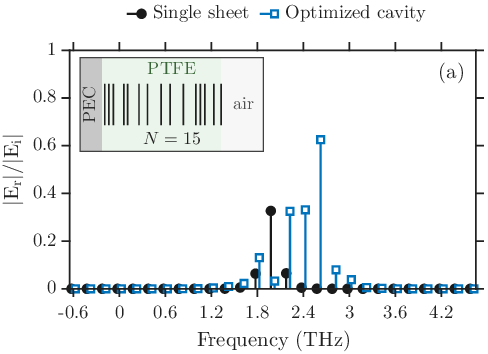}
    \caption{Independent gaps $d_k$ optimized using the target-only objective \eqref{eq39}.}
    \label{fig6a}
  \end{subfigure}\hfill
  \begin{subfigure}[t]{0.48\textwidth}
    \centering
    \includegraphics[width=\linewidth]{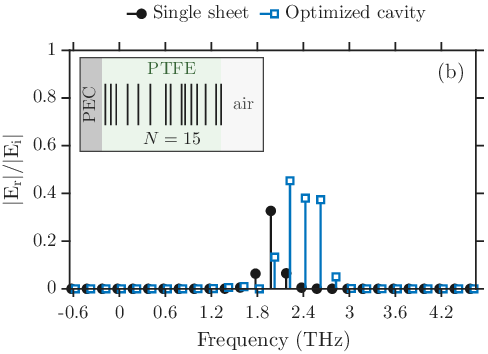}
    \caption{Independent gaps $d_k$ optimized using the multi-objective formulation \eqref{eq40} with $w=1$.}
    \label{fig6b}
  \end{subfigure}

  \caption{Reflected spectra for the time-modulated graphene cavity targeting the harmonic at $f_{+3}=2.6~\mathrm{THz}$. Panel (a) shows the baseline target-only optimization using \eqref{eq39}, whereas panel (b) shows the multi-objective optimization using \eqref{eq40} with $w=1$.}
  \label{fig6}
\end{figure*}

At this point, to quantify the spectral isolation achieved by the two optimization strategies, we introduce the following two metrics,
\begin{subequations}\label{eq41}
\begin{align}
S_{\max} &= \frac{|E_{\rm r}^{(t)}|}{\max_{n\neq t}\Big\{|E_{\rm r}^{(n)}|\Big\}},\label{eq41a}\\
S_{\Sigma} &= \frac{|E_{\rm r}^{(t)}|}{\sum_{n\neq t}|E_{\rm r}^{(n)}|},\label{eq41b}
\end{align}
\end{subequations}
where $E_{\rm r}^{(t)}$ denotes the reflected amplitude at the targeted harmonic. Both metrics are dimensionless, and larger values indicate stronger spectral isolation of the target harmonic. The first metric measures the dominance of the target harmonic over the strongest of the other harmonics, whereas the second compares the target to the total non-target spectral "leakage".

We then apply the multi-objective formulation of \eqref{eq40}, again under the same practical bounds $0.2\lambda_0 \le d_k \le \lambda_0$, and perform a parametric study over the weighting parameter $w$. Among the values examined, $w=1$ was found to provide the most favorable trade-off. In this case, the reflected amplitude at the targeted harmonic reaches $|E_{\rm r}|=0.468$, corresponding to an enhancement of $59.56\,$dB relative to the single-sheet response. Although the targeted harmonic amplitude is lower than that obtained in the baseline optimization, the overall undesirable harmonic content is reduced substantially, with the total non-target leakage decreasing from $0.982$ to $0.557$. 
A direct comparison between the two designs shows that the single-objective optimization retains the larger target amplitude, whereas the multi-objective formulation of \eqref{eq40} produces a ``cleaner'' overall spectrum. Specifically, the $S_{\max}$ metric of \eqref{eq41} changes only marginally from $1.8886$ for the baseline design to $1.9582$ for the multi-objective design, indicating that the strongest competing sideband remains of comparable magnitude in both cases. In contrast, the total-leakage metric of \eqref{eq41} improves more noticeably, increasing from $0.6367$ in the baseline case to $0.8405$ for the multi-objective design. Therefore, under the imposed realistic cavity bounds, the principal effect of \eqref{eq40} is to suppress the aggregate non-target spectral content and thereby improve the overall spectral cleanliness of the reflected response.

Table~\ref{table1} lists the optimized cavity gaps for both strategies. The comparison shows that both designs exploit the full range of cavity spacings, but converge to distinct non-uniform spacing configurations, reflecting the different optimization properties encoded in \eqref{eq39} and \eqref{eq40}. Figure~\ref{fig6} compares the corresponding reflected spectra against the single-sheet reference. The design corresponding to the left column of Table~\ref{table1} yields the strongest amplification at the target harmonic, but retains substantial energy in adjacent sidebands, as depicted in Fig.~\ref{fig6a}. In contrast, as illustrated in Fig.~\ref{fig6b}, the multi-objective design using the spacing dimensions of the right column of Table~\ref{table1} exhibits a reduced target amplitude, yet a noticeably lower total non-target leakage, and consequently a cleaner overall reflected spectrum. These results demonstrate that harmonic engineering is inherently a trade-off between peak enhancement and global spectral suppression. They also confirm that the Taylor-expanded conductivity model is essential for accurately capturing and optimizing third and higher generated harmonics for small values of $\mu_{{\rm c},0}$.

\begin{table*}[t]
\centering
\small
\caption{Optimized cavity gaps for the $N=15$ graphene cavity with PTFE filling, targeting the third generated harmonic at $f_{+3}=2.6$~THz. The table compares the baseline target-only optimization using \eqref{eq39} with the multi-objective optimization using \eqref{eq40} with $w=1$. The gaps are reported in millimeters and normalized to the free-space wavelength $\lambda_0$ at $f_0=2$~THz. In both cases, the design bounds are $0.2\lambda_0 \le d_k \le \lambda_0$.}
\label{table1}

\setlength{\tabcolsep}{6pt}
\begin{tabular}{
@{}
c r r @{\hspace{0.8em}}
r r
@{}
}
\toprule
& \multicolumn{2}{c}{\shortstack[c]{Target-only\\[2pt] Eq.~(39)}}
& \multicolumn{2}{c}{\shortstack[c]{Multi-objective\\[2pt] Eq.~(40), $w=1$}} \\
\cmidrule(lr){2-3}\cmidrule(lr){4-5}
\multicolumn{1}{c}{\rule{0pt}{2.6ex}Sheet index $k$}
& \multicolumn{1}{c}{$d_k$ (mm)}
& \multicolumn{1}{c}{$d_k/\lambda_0$}
& \multicolumn{1}{c}{$d_k$ (mm)}
& \multicolumn{1}{c}{$d_k/\lambda_0$} \\
\midrule
1  & 0.0802 & 0.5350 & 0.0471 & 0.3140 \\
2  & 0.1000 & 0.6669 & 0.0960 & 0.6403 \\
3  & 0.0499 & 0.3326 & 0.0769 & 0.5133 \\
4  & 0.0497 & 0.3316 & 0.0543 & 0.3623 \\
5  & 0.1380 & 0.9205 & 0.0556 & 0.3707 \\
6  & 0.1455 & 0.9705 & 0.0334 & 0.2226 \\
7  & 0.1009 & 0.6734 & 0.1001 & 0.6681 \\
8  & 0.1499 & 1.0000 & 0.0430 & 0.2872 \\
9  & 0.0941 & 0.6275 & 0.1424 & 0.9501 \\
10 & 0.1255 & 0.8370 & 0.1108 & 0.7391 \\
11 & 0.0417 & 0.2779 & 0.0988 & 0.6593 \\
12 & 0.1165 & 0.7775 & 0.1033 & 0.6892 \\
13 & 0.0502 & 0.3349 & 0.0501 & 0.3344 \\
14 & 0.0444 & 0.2965 & 0.0500 & 0.3337 \\
15 & 0.0300 & 0.2001 & 0.0303 & 0.2021 \\
\bottomrule
\end{tabular}
\end{table*}

\subsection{3.3 Symmetry-Induced Enhancement of the Upper Second-Order Even Harmonic at Zero Bias}
Let us now consider a third design regime that directly exploits the symmetry property established in \textit{Section 2.2}. In this example, the goal is to selectively enhance the upper second-order even harmonic, namely $f_{+2}=f_0+2f_{\rm mod}=2.4~\mathrm{THz}$. Note that the even symmetry of the surface conductivity function with respect to the chemical potential is exploited to generate a spectrum with exclusively even harmonics; therefore, the constant chemical potential is set to zero, unlike the previous examples, while the modulated part is set to $\Delta\mu=0.07~\mathrm{eV}$ leading in $\mu_{\rm c}(t)=\Delta\mu \cos(\omega_{\rm mod} t)$. Note that at room's temperature, the selected modulation excursion remains within the local convergence disk of the Taylor-expanded Drude prefactor, which is obtained $R(0)=\pi k_{\rm B}T \approx 0.08072~\mathrm{eV}$ for $\mu_{{\rm c},0}=0$ (see \textit{Appendix B}). In terms of the remaining attributes of the device, a total number of $N=15$ time-modulated graphene sheets are utilized with polytetrafluoroethylene (PTFE) dielectric spacers of $\varepsilon_{\rm r}=2$.

\begin{table*}[t]
\centering
\small
\caption{Optimized cavity gaps for the $N=15$ graphene cavity with PTFE filling, targeting the upper second-order even harmonic at $f_{+2}=2.4$~THz. The design is obtained using the target-only objective in \eqref{eq39}. The gaps are reported in millimeters and normalized to the free-space wavelength $\lambda_0$ at $f_0=2$~THz. The design bounds are $0.2\lambda_0 \le d_k \le \lambda_0$.}\label{table2}
\setlength{\tabcolsep}{8pt}
\begin{tabular}{
c c c
}
\toprule
\multicolumn{1}{c}{\rule{0pt}{2.6ex}Sheet index $k$}
& \multicolumn{1}{c}{$d_k$ (mm)}
& \multicolumn{1}{c}{$d_k/\lambda_0$} \\
\midrule
1  & 0.1411 & 0.9414 \\
2  & 0.1109 & 0.7396 \\
3  & 0.1446 & 0.9648 \\
4  & 0.0451 & 0.3009 \\
5  & 0.1353 & 0.9028 \\
6  & 0.0509 & 0.3396 \\
7  & 0.1045 & 0.6968 \\
8  & 0.0528 & 0.3524 \\
9  & 0.1436 & 0.9580 \\
10 & 0.0923 & 0.6158 \\
11 & 0.0495 & 0.3305 \\
12 & 0.0971 & 0.6481 \\
13 & 0.0517 & 0.3446 \\
14 & 0.0477 & 0.3184 \\
15 & 0.0990 & 0.6603 \\
\bottomrule
\end{tabular}
\end{table*}

\begin{figure}[t]
  \centering
  \includegraphics[width=0.5\linewidth]{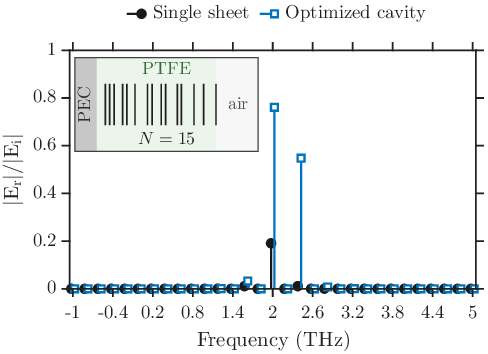}
  \caption{Reflected spectra from the optimized cavity versus the single-sheet reference, for the time-modulated graphene cavity targeting the $(n_0+2)$-th harmonic at $f_{+2}=2.4$~THz, with $\mu_{{\rm c},0}=0$. }\label{fig7}
\end{figure}

As in the previous subsection, the reflected spectrum is written as $\mathbf{E}_{\rm r}(\mathbf{d})=\mathbf{R}_{\rm tot}(\mathbf{d})\,\mathbf{e}_{n,0}$, where $\mathbf{d}=\{d_1,d_2,\dots,d_{15}\}$ denotes the set of cavity gaps, which are again allowed to vary independently. Since the purpose of the present example is to demonstrate the even-harmonic response arising from zero-centered modulation, the design is obtained here by maximizing the target harmonic through the same objective already introduced in \eqref{eq39}, namely $J(\mathbf{d})=-|E_{\rm r}^{(n_t)}(\mathbf{d})|$, where $n_t=(n_0)+2$ denotes the target harmonic index, i.e., the upper second-order harmonic. The optimization is performed under the practical bounds $0.2\lambda_0 \le d_k \le \lambda_0$, for $k=1,2,\ldots,15$. 
%
The optimized cavity yields a reflected amplitude of $|E_{\rm r}|=0.5477$ at the target harmonic $f_{+2}=2.4~\mathrm{THz}$, starting from a single-sheet reference value of $|E_{\rm r}|=0.01123$. This corresponds to an enhancement factor of $48.78$, or an amplitude gain of $33.76~\mathrm{dB}$. The optimized cavity gaps are listed in Table~\ref{table2}, while the corresponding reflected spectrum is shown in Fig. \ref{fig7}. It is observed that the optimized cavity strongly reinforces the $(n_0+2)$-th component, while preserving the strictly even character predicted by the theory.

Because odd harmonics are symmetry-forbidden in the centered-bias case, it is more meaningful here to assess spectral isolation within the even-harmonic family. To this end, we introduce the even-subspace counterparts of the selectivity metrics of \eqref{eq41}, namely 

\begin{subequations}\label{eq42}
\begin{align}
S_{\max}^{(\mathrm{even})} &= \frac{|E_{\rm r}^{(t)}|}{\max_{\substack{n\in\mathcal{E}\\ n\neq t}} |E_{\rm r}^{(n)}|},\label{eq42a}\\
S_{\Sigma}^{(\mathrm{even})} &= \frac{|E_{\rm r}^{(t)}|}{\sum_{\substack{n\in\mathcal{E}\\ n\neq t}} |E_{\rm r}^{(n)}|},\label{eq42b}
\end{align}
\end{subequations}

\noindent where $\mathcal{E}$ denotes the set of even harmonics and $t$ the targeted harmonic index. For the optimized design, one obtains $S_{\max}^{(\mathrm{even})}=0.7199$ and $S_{\Sigma}^{(\mathrm{even})}=0.6822$. These values show that the residual competition is dominated almost entirely by the carrier component at $f_0=2~\mathrm{THz}$, whose reflected amplitude reaches $|E_{\rm r}|=0.7608$. Nevertheless, all odd harmonics vanish to numerical precision, and the remaining higher-order even sidebands are substantially weaker than the target response. In particular, the $(n_0+4)$-th harmonic remains at $|E_{\rm r}|=7.49\times10^{-3}$, while the higher even orders decay rapidly.

Therefore, the present example confirms a qualitatively different harmonic-engineering regime from those examined in \textit{Sections 3.1} and \textit{3.2}. Here, the main effect is not the suppression of a broad set of unwanted harmonics, but the symmetry-enforced elimination of the odd ones together with the cavity-assisted enhancement of a selected even frequency. Although the central frequency remains stronger than the targeted $(n_0+2)$-th harmonic, the reflected response is confined strictly to the even-harmonic subspace, which makes the component at $f_0+2f_{\rm mod}$ particularly attractive for practical frequency-conversion schemes. In a system-level implementation, the residual central frequency could be further reduced by downstream spectral filtering, whereas the cavity itself already enforces the more fundamental even-harmonic selection rule.

\section{4. Conclusion}

In this work, we developed a semi-analytic framework for modeling and designing dispersive, time-modulated graphene structures. Starting from the boundary conditions of a conductive, time-dependent, and dispersive interface, we derived a frequency-domain operator formulation in which temporal modulation appears explicitly as spectral coupling among generated harmonics. Discretization on a uniform frequency grid yields a compact matrix system that serves as a building block for multilayer and resonant configurations.

For graphene in the intraband regime, we incorporated two conductivity models, namely a Taylor-expanded Drude formulation that rigorously captures higher-order temporal mixing, and a linearized high-bias approximation that reduces coupling to nearest-neighbor harmonics. Comparison with full-wave FDTD simulations for a time-varying graphene stack terminated by a PEC wall validated the transfer-matrix implementation and clarified the regimes of validity. The Taylor model matches the FDTD across all tested chemical potentials, while the approximation becomes accurate only at sufficiently high bias. In addition, the Taylor-expanded formulation captures symmetry-induced effects that lie outside the range of validity of the linearized model, including the purely even harmonic generation that arises under zero-centered modulation.

We then leveraged the framework for optimization-driven harmonic engineering. In the high-bias regime, the reduced model enabled selective enhancement of the upper and lower first-order harmonics through optimization of the cavity gaps, while a parametric study with respect to the number of graphene sheets showed that most of the available first-order reinforcement is already captured once a short multilayer cavity is formed. In a second design case, targeting the third generated harmonic at a lower chemical-potential value, the Taylor model combined with optimization over independent cavity gaps showed that harmonic engineering in this regime is governed by a trade-off between peak enhancement and overall spectral suppression. The target-only formulation yielded the largest target response, whereas the multi-objective formulation reduced the total non-target leakage and produced a cleaner reflected spectrum. Finally, in a zero-bias modulation regime, we showed that the symmetry of the Drude prefactor can be exploited directly to realize cavity-assisted enhancement of an upper second-order even harmonic, with odd harmonics suppressed to numerical precision.

Overall, the present results establish a rigorous and versatile modeling and design framework for dispersive time-modulated graphene cavities. They show that the same formalism can describe first harmonic reinforcement in the high-bias regime, higher-order harmonic engineering beyond the linear regime, and symmetry-induced generation of even harmonics under zero-centered modulation. In this sense, the proposed framework offers a scalable route to compact frequency-conversion structures based on time-modulated graphene or other quasi-surface media for the mm-wave, THz and optical spectra.

\appendix

\section{Appendix A: Linearized Approximation for the Graphene Conductivity}

For completeness, we summarize here the linearized approximation for the temporally modulated graphene conductivity, which is accurate for high-bias scenarios. This model has been employed in earlier works~\cite{Salary2018PRB, Sedeh2022TAP, Karamanos2024META} and is recovered naturally as a limiting case of the general Taylor-expanded formulation presented in the main text.

When the chemical potential remains sufficiently large compared to the thermal energy, the logarithmic term in the Drude prefactor becomes negligible, and \eqref{eq8} reduces to the approximate form
\begin{equation}\label{eq43}
A(\mu_{\rm c})\simeq \frac{e^2\mu_{\rm c}}{\pi\hbar^2}.
\end{equation}
Under this condition, the conductivity becomes effectively linear in the chemical potential, and higher-order modulation effects can be neglected.

Assuming the temporal modulation
\begin{equation}\label{eq44}
\mu_{\rm c}(t)=\mu_{{\rm c},0}+\Delta\mu\cos\theta,
\end{equation}
with $\theta=\omega_{\mathrm{mod}}t$, and using the modulation-depth definition
\begin{equation}\label{eq45}
\Delta\mu=\mu_{{\rm c},0}M,
\end{equation}
the conductivity modulation corresponds to retaining only the first-order term of the Taylor expansion. Substituting this approximation into the time-domain response \eqref{eq10} and transforming to the frequency domain yields
\begin{equation}\label{eq46}
\begin{aligned}
\sigma(\omega)
=
\frac{A_{\mu_{{\rm c},0}}}{j\omega + 2\Gamma}
\,\ast\,
\Big[
\delta(\omega)
+ \frac{M}{2}\delta(\omega-\omega_{\mathrm{mod}})
+ \frac{M}{2}\delta(\omega+\omega_{\mathrm{mod}})
\Big].
\end{aligned}
\end{equation}

This expression shows that, in the high-bias regime, the temporally modulated conductivity generates only the first pair of modulation sidebands at $\omega\pm\omega_{\mathrm{mod}}$. Within the operator formulation of \textit{Section}~2.1, this implies that the frequency-coupling matrix $\mathbf{S}$ becomes tridiagonal, with nonzero entries only on the main diagonal and the two adjacent off-diagonals. Physically, this corresponds to nearest-neighbor coupling between harmonics, which is the direct consequence of truncating the conductivity expansion at first order.

The linearized approximation, therefore, provides a compact and computationally efficient model for sufficiently large bias levels and moderate modulation depths. At the same time, phenomena such as higher harmonic generation, symmetry-induced effects, and strongly nonlinear modulation regimes lie outside its range of validity and require the full Taylor-expanded formulation developed in the main text.

\section{Appendix B: Radius of Convergence of the Taylor-Expanded Drude Prefactor}
We summarize herein the complex analysis argument underlying the validity range of the Taylor-expanded conductivity model used in \textit{Section}~2.2. Let $f(z)$ be an analytic function expanded about a point $z_0\in\mathbb{C}$ as
\begin{equation}\label{eq47}
    f(z)=\sum_{n=0}^{\infty} a_n (z-z_0)^n.
\end{equation}
A standard result of complex analysis states that this power series converges inside a disk $|z-z_0|<\rho$, where $\rho$ is the radius of convergence, and that $\rho$ is determined by the distance from $z_0$ to the nearest singularity of $f(z)$ \cite{OlverComplexAnalysis}. In particular, the convergence is local, even when the original function is perfectly well-defined on the real axis.

In the present work, the relevant quantity is the finite-temperature Drude prefactor
\begin{equation}\label{eq48}
A(\mu)=
\frac{e^2}{\pi\hbar^2}
\left[
\mu+2k_{\rm B} T\ln\!\left(1+e^{-\mu/(k_{\rm B} T)}\right)
\right].
\end{equation}
The only non-entire contribution in \eqref{eq48} is the logarithmic term. Its singularities arise when the argument of the logarithm vanishes, namely when 
\begin{equation}\label{eq49}
1+e^{-\mu/(k_{\rm B} T)}=0.
\end{equation}
Solving \eqref{eq49} gives
\begin{equation}\label{eq50}
e^{-\mu/(k_{\rm B} T)}=-1=e^{j(2\ell+1)\pi},
\qquad \ell\in\mathbb{Z},
\end{equation}
and therefore
\begin{equation}\label{eq51}
\mu=j(2\ell+1)\pi k_{\rm B} T,\qquad \ell\in\mathbb{Z}.
\end{equation}
Hence, for a real expansion point $\mu_{{\rm c},0}$, the nearest singularities are located at $\mu=\pm i\pi k_{\rm B} T$. Their distance from $\mu_{{\rm c},0}$ is
\begin{equation}\label{eq52}
R(\mu_{{\rm c},0})
=
\left|\mu_{{\rm c},0}-j\pi k_{\rm B} T\right|
=
\sqrt{\mu_{{\rm c},0}^{\,2}+(\pi k_{\rm B} T)^2}.
\end{equation}
This quantity defines the convergence radius of the Taylor expansion of $A(\mu)$ about $\mu_{{\rm c},0}$. Consequently, if the temporal modulation
\begin{equation}\label{eq53}
\mu_{\rm c}(t)=\mu_{{\rm c},0}+\Delta\mu\cos(\omega_{\rm mod} t)
\end{equation}
satisfies
\begin{equation}\label{eq54}
\Delta\mu<R(\mu_{{\rm c},0}),
\end{equation}
then the full modulation trajectory remains inside the convergence disk of the local Taylor representation. On the other hand, when $\Delta\mu$ exceeds this radius, the truncated Taylor series is no longer guaranteed to converge uniformly along the full modulation path, even though the exact prefactor \eqref{eq48} remains well-defined for real values of $\mu_{\rm c}(t)$.

The preceding observation elucidates a pivotal modeling point, namely that the limitation is associated with the employment of a local Taylor expansion in the chemical potential. For this reason, strongly modulated regimes may require either a sufficiently restricted parameter range or an alternative harmonic extraction strategy based directly on the exact time-varying prefactor.

\section{Appendix C: Recursive Multiple-Scattering Formulation}

We present here an alternative derivation of the multilayer response based on a recursive multiple-scattering formulation. This approach provides a complementary physical interpretation of the structure analyzed in \textit{Section}~2.3, where the main text adopts the transfer-matrix method for numerical implementation.

From a physical standpoint, the response of the multilayer system can be interpreted as the coherent superposition of an infinite sequence of internal reflections between the graphene sheets and the terminating PEC wall. To formalize this picture, we introduce the diagonal propagation operator across the $k$-th dielectric spacer
\begin{equation}\label{eq55}
\mathbf{D}_k = \mathrm{diag}\!\left\{ e^{-j k_n^{(k)} d_k} \right\},
\end{equation}
where $k_n^{(k)}=\omega_n\sqrt{\mu_k\varepsilon_k}$ denotes the wavenumber of the $n$-th harmonic in the $k$-th spacer medium.

In the general case, the media on the two sides of a graphene sheet may be different, so that the scattering operators are direction-dependent. Let $\mathbf{R}_{{\rm g},k}^{\mathrm{L}}$ and $\mathbf{R}_{{\rm g},k}^{\mathrm{R}}$ denote the reflection operators for incidence from the left and right sides of the $k$-th sheet, respectively, and let $\mathbf{T}_{g,k}^{\mathrm{L}\rightarrow\mathrm{R}}$ and $\mathbf{T}_{{\rm g},k}^{\mathrm{R}\rightarrow\mathrm{L}}$ denote the corresponding transmission operators. We define the effective reflection operator $\mathbf{R}_j^{\mathrm{eff}}$ as the reflection seen at the plane immediately to the left of the $j$-th graphene sheet, looking into the portion of the structure extending from that sheet to the terminating PEC wall, so that
\begin{equation}\label{eq56}
\mathbf{E}_{j-1}^-=\mathbf{R}_j^{\mathrm{eff}}\,\mathbf{E}_{j-1}^+.
\end{equation}

Starting from the terminating PEC, one first considers the subsystem formed by the $N$-th graphene sheet, the $N$-th dielectric spacer, and the PEC wall. At the PEC termination, the tangential electric field must vanish, which implies that the backward-propagating field at the right side of the $N$-th sheet is related to the forward-propagating one through
\begin{equation}\label{eq57}
\mathbf{E}_N^- = -\,\mathbf{D}_N^2\,\mathbf{E}_N^+.
\end{equation}
Substituting this relation into the directional scattering equations of the $N$-th sheet yields the effective reflection operator
\begin{equation}\label{eq58}
\mathbf{R}_{N}^{\mathrm{eff}}
=
\mathbf{R}_{{\rm g},N}^{\mathrm{L}}
-
\mathbf{T}_{{\rm g},N}^{\mathrm{R}\rightarrow\mathrm{L}}
\mathbf{D}_N^2
\bigl(
\mathbf{I}
+
\mathbf{R}_{{\rm g},N}^{\mathrm{R}}\mathbf{D}_N^2
\bigr)^{-1}
\mathbf{T}_{{\rm g},N}^{\mathrm{L}\rightarrow\mathrm{R}},
\end{equation}
which accounts for all multiple round-trip interactions between the $N$-th sheet and the PEC boundary.

Proceeding recursively toward the input side, let
\begin{equation}\label{eq59}
\mathbf{X}_j
=
\mathbf{D}_j\,\mathbf{R}_{j+1}^{\mathrm{eff}}\,\mathbf{D}_j
\end{equation}
denote the effective round-trip reflection operator of the subsystem located to the right of the $j$-th graphene sheet, referred back to the plane immediately to the right of that sheet. The effective reflection operator seen at the input side of the $j$-th graphene sheet then becomes
\begin{equation}\label{eq60}
\mathbf{R}_{j}^{\mathrm{eff}}
=
\mathbf{R}_{{\rm g},j}^{\mathrm{L}}
+
\mathbf{T}_{{\rm g},j}^{\mathrm{R}\rightarrow\mathrm{L}}
\mathbf{X}_i
\bigl(
\mathbf{I}
-
\mathbf{R}_{{\rm g},j}^{\mathrm{R}}\mathbf{X}_j
\bigr)^{-1}
\mathbf{T}_{{\rm g},j}^{\mathrm{L}\rightarrow\mathrm{R}},
\qquad j=1,2,\dots,N-1.
\end{equation}
Equivalently, substituting the definition of $\mathbf{X}_j$ gives
\begin{equation}\label{eq61}
\mathbf{R}_{j}^{\mathrm{eff}}
=
\mathbf{R}_{{\rm g},j}^{\mathrm{L}}
+
\mathbf{T}_{{\rm g},j}^{\mathrm{R}\rightarrow\mathrm{L}}
\mathbf{D}_j\,\mathbf{R}_{j+1}^{\mathrm{eff}}\,\mathbf{D}_j
\bigl(
\mathbf{I}
-
\mathbf{R}_{{\rm g},j}^{\mathrm{R}}
\mathbf{D}_j\,\mathbf{R}_{j+1}^{\mathrm{eff}}\,\mathbf{D}_j
\bigr)^{-1}
\mathbf{T}_{{\rm g},j}^{\mathrm{L}\rightarrow\mathrm{R}}.
\end{equation}

Once the recursion reaches the first graphene sheet, the reflected field at the input plane follows directly as
\begin{equation}\label{eq62}
\mathbf{E}_{\rm r} = \mathbf{R}_{1}^{\mathrm{eff}}\,\mathbf{E}_{\rm i} .
\end{equation}

In the symmetric case where the media on the two sides of a graphene sheet are identical, the directional operators reduce to
\begin{equation}\label{eq63}
\mathbf{R}_{{\rm g},k}^{\mathrm{L}}=\mathbf{R}_{{\rm g},k}^{\mathrm{R}}\equiv \mathbf{R}_{{\rm g},k},
\qquad
\mathbf{T}_{{\rm g},k}^{\mathrm{L}\rightarrow\mathrm{R}}
=
\mathbf{T}_{{\rm g},k}^{\mathrm{R}\rightarrow\mathrm{L}}
\equiv \mathbf{T}_{{\rm g},k}.
\end{equation}
In that case, \eqref{eq58} becomes
\begin{equation}\label{eq64}
\mathbf{R}_{N}^{\mathrm{eff}}
=
\mathbf{R}_{{\rm g},N}
-
\mathbf{T}_{{\rm g},N}\mathbf{D}_N^2
\bigl(
\mathbf{I}
+
\mathbf{R}_{{\rm g},N}\mathbf{D}_N^2
\bigr)^{-1}
\mathbf{T}_{{\rm g},N},
\end{equation}
while the recursive step reduces to
\begin{equation}\label{eq65}
\mathbf{R}_{j}^{\mathrm{eff}}
=
\mathbf{R}_{{\rm g},j}
+
\mathbf{T}_{{\rm g},j}\mathbf{D}_j\,\mathbf{R}_{j+1}^{\mathrm{eff}}\,\mathbf{D}_j
\bigl(
\mathbf{I}
-
\mathbf{R}_{{\rm g},j}\mathbf{D}_j\,\mathbf{R}_{j+1}^{\mathrm{eff}}\,\mathbf{D}_j
\bigr)^{-1}
\mathbf{T}_{{\rm g},j}.
\end{equation}
In the cavity configurations considered in the main text, this symmetric reduction applies to the interior graphene sheets when the cavity medium is uniform, whereas the exterior-facing sheet generally requires the full directional formulation if the exterior and cavity media differ.

This recursive formulation highlights the physical mechanism underlying the multilayer response, namely the coherent accumulation of frequency-coupled round trips within the cavity. It provides additional physical insight into the role of internal reflections and harmonic coupling in time-modulated graphene resonators, and is fully consistent with the transfer-matrix treatment adopted in the main text.

\printbibliography

@article{Koutzoglou2026,
  author  = {Koutzoglou, I. and Amanatiadis, S. and Kantartzis, N. V.},
  title   = {Robust and Integrable Time-Varying Metamaterials: A Systematic Survey and Coherent Mapping},
  journal = {Nanomaterials},
  year    = {2026},
  volume  = {16},
  pages   = {195},
  doi     = {10.3390/nano16030195}
}

@article{galiffi2022photonics,
  title={Photonics of time-varying media},
  author={Galiffi, Emanuele and Tirole, Romain and Yin, Shixiong and Li, Huanan and Vezzoli, Stefano and Huidobro, Paloma A and Silveirinha, M{\'a}rio G and Sapienza, Riccardo and Al{\`u}, Andrea and Pendry, John B},
  journal={Advanced Photonics},
  volume={4},
  number={1},
  pages={014002--014002},
  year={2022},
  publisher={Society of Photo-Optical Instrumentation Engineers}
}

@article{amanatiadis2025enhanced,
  title={Enhanced mm-wave frequency up-conversion via a time-varying graphene aperture on a cavity resonator},
  author={Amanatiadis, Stamatios and Karamanos, Theodosios and Lemoult, Fabrice and Kantartzis, Nikolaos V},
  journal={Micromachines},
  volume={16},
  number={6},
  pages={679},
  year={2025},
  publisher={MDPI}
}

@manual{GlobalOptimizationToolbox,
  author = {{The MathWorks Inc.}},
  title  = {Global Optimization Toolbox},
  year   = {2026},
  address = {Natick, Massachusetts, United States},
  note   = {Version R2026a},
  url    = {https://www.mathworks.com/products/global-optimization.html}
}

@article{caloz2019spacetime,
  title={Spacetime metamaterials—part I: general concepts},
  author={Caloz, Christophe and Deck-L{\'e}ger, Zo{\'e}-Lise},
  journal={IEEE Transactions on Antennas and Propagation},
  volume={68},
  number={3},
  pages={1569--1582},
  year={2019},
  publisher={IEEE}
}

@article{movahediqomi2026stacked,
  title={Stacked Time-Varying Metasurfaces},
  author={Movahediqomi, Mostafa and Tretyakov, Sergei and Asadchy, Viktar and Wang, Xuchen},
  journal={Advanced Optical Materials},
  pages={e02831},
  year={2026},
  publisher={Wiley Online Library}
}

@article{asgari2024theory,
  title={Theory and applications of photonic time crystals: a tutorial},
  author={Asgari, Mohammad M and Garg, Puneet and Wang, Xuchen and Mirmoosa, Mohammad S and Rockstuhl, Carsten and Asadchy, Viktar},
  journal={Advances in optics and photonics},
  volume={16},
  number={4},
  pages={958--1063},
  year={2024},
  publisher={Optica Publishing Group}
}

@article{wang2025expanding,
  title={Expanding momentum bandgaps in photonic time crystals through resonances},
  author={Wang, X and Garg, P and Mirmoosa, MS and Lamprianidis, AG and Rockstuhl, C and Asadchy, VS},
  journal={Nature Photonics},
  volume={19},
  number={2},
  pages={149--155},
  year={2025},
  publisher={Nature Publishing Group UK London}
}

@article{altares2017frequency,
  title={Frequency comb generation using plasmonic resonances in a time-dependent graphene ribbon array},
  author={Altares Menendez, Galaad and Maes, Bjorn},
  journal={Physical Review B},
  volume={95},
  number={14},
  pages={144307},
  year={2017},
  publisher={APS}
}

@article{asadchy2022parametric,
  title={Parametric Mie resonances and directional amplification in time-modulated scatterers},
  author={Asadchy, V and Lamprianidis, AG and Ptitcyn, G and Albooyeh, M and Rituraj and Karamanos, T and Alaee, R and Tretyakov, SA and Rockstuhl, C and Fan, S},
  journal={Physical Review Applied},
  volume={18},
  number={5},
  pages={054065},
  year={2022},
  publisher={APS}
}

@article{Liu2016ACSPhotonics,
  author  = {Liu, Zizhuo and Li, Zhongyang and Aydin, Koray},
  title   = {Time-Varying Metasurfaces Based on Graphene Microribbon Arrays},
  journal = {ACS Photonics},
  year    = {2016},
  volume  = {3},
  number  = {11},
  pages   = {2035--2039},
  doi     = {10.1021/acsphotonics.6b00653}
}

@article{Zhou2020,
  author  = {Zhou, Yiyu and Alam, M. Zahirul and Karimi, Mohammad and Upham, Jeremy and Reshef, Orad and Liu, Cong and Willner, Alan E. and Boyd, Robert W.},
  title   = {Broadband Frequency Translation through Time Refraction in an Epsilon-Near-Zero Material},
  journal = {Nature Communications},
  year    = {2020},
  volume  = {11},
  pages   = {2180},
  doi     = {10.1038/s41467-020-15682-2}
}

@article{CorreasSerrano2016,
  author  = {Correas-Serrano, David and Gomez-Diaz, Juan S. and Sounas, Dimitrios L. and Hadad, Yakir and Alvarez-Melcon, Alejandro and Alu, Andrea},
  title   = {Nonreciprocal Graphene Devices and Antennas Based on Spatiotemporal Modulation},
  journal = {IEEE Antennas and Wireless Propagation Letters},
  year    = {2016},
  volume  = {15},
  pages   = {1529--1532},
  doi     = {10.1109/LAWP.2015.2510818}
}

@article{Wang2020,
  author  = {Wang, Xuchen and Diaz-Rubio, Ana and Li, Huanan and Tretyakov, Sergei A. and Alu, Andrea},
  title   = {Theory and Design of Multifunctional Space-Time Metasurfaces},
  journal = {Physical Review Applied},
  year    = {2020},
  volume  = {13},
  number  = {4},
  pages   = {044040},
  doi     = {10.1103/PhysRevApplied.13.044040}
}

@article{Yu2009,
  author  = {Yu, Zongfu and Fan, Shanhui},
  title   = {Complete Optical Isolation Created by Indirect Interband Photonic Transitions},
  journal = {Nature Photonics},
  year    = {2009},
  volume  = {3},
  pages   = {91--94},
  doi     = {10.1038/nphoton.2008.273}
}

@article{Shi2017ACSPhotonics,
  author  = {Shi, Yu and Han, Seunghoon and Fan, Shanhui},
  title   = {Optical Circulation and Isolation Based on Indirect Photonic Transitions of Guided Resonance Modes},
  journal = {ACS Photonics},
  year    = {2017},
  volume  = {4},
  number  = {7},
  pages   = {1639--1645},
  doi     = {10.1021/acsphotonics.7b00420}
}

@article{Lyubarov2022,
  author  = {Lyubarov, Mark and others},
  title   = {Amplified Emission and Lasing in Photonic Time Crystals},
  journal = {Science},
  year    = {2022},
  volume  = {377},
  pages   = {425--428},
  doi     = {10.1126/science.abo3324}
}

@article{Lee2021,
  author  = {Lee, Seojoo and Park, Jagang and Cho, Hyukjoon and Wang, Yifan and Kim, Brian and Daraio, Chiara and Min, Bumki},
  title   = {Parametric Oscillation of Electromagnetic Waves in Momentum Band Gaps of a Spatiotemporal Crystal},
  journal = {Photonics Research},
  year    = {2021},
  volume  = {9},
  pages   = {142--150}
}

@article{PachecoPena2020Optica,
  author  = {Pacheco-Pe{\~n}a, Victor and Engheta, Nader},
  title   = {Antireflection Temporal Coatings},
  journal = {Optica},
  year    = {2020},
  volume  = {7},
  pages   = {323--331}
}

@article{Li2021,
  author  = {Li, Huanan and Al{\`u}, Andrea},
  title   = {Temporal Switching to Extend the Bandwidth of Thin Absorbers},
  journal = {Optica},
  year    = {2021},
  volume  = {8},
  pages   = {24--29}
}

@article{Castaldi2021,
  author  = {Castaldi, Giuseppe and Pacheco-Pe{\~n}a, Victor and Moccia, Massimo and Engheta, Nader and Galdi, Vincenzo},
  title   = {Exploiting Space-Time Duality in the Synthesis of Impedance Transformers via Temporal Metamaterials},
  journal = {Nanophotonics},
  year    = {2021},
  volume  = {10},
  number  = {14},
  pages   = {3687--3699},
  doi     = {10.1515/nanoph-2021-0231}
}

@article{PachecoPena2020LSA,
  author  = {Pacheco-Pe{\~n}a, Victor and Engheta, Nader},
  title   = {Temporal Aiming},
  journal = {Light: Science \& Applications},
  year    = {2020},
  volume  = {9},
  pages   = {129},
  doi     = {10.1038/s41377-020-00360-1}
}

@article{Ramaccia2017PRB,
  author  = {Ramaccia, Davide and Sounas, Dimitrios L. and Al{\`u}, Andrea and Toscano, Alessandro and Bilotti, Filiberto},
  title   = {Doppler Cloak Restores Invisibility to Objects in Relativistic Motion},
  journal = {Physical Review B},
  year    = {2017},
  volume  = {95},
  number  = {7},
  pages   = {075113},
  doi     = {10.1103/PhysRevB.95.075113}
}

@article{Fridman2012Nature,
  author  = {Fridman, Moti and Farsi, Alireza and Okawachi, Yoshitomo and Gaeta, Alexander L.},
  title   = {Demonstration of Temporal Cloaking},
  journal = {Nature},
  year    = {2012},
  volume  = {481},
  pages   = {62--65},
  doi     = {10.1038/nature10695}
}

@article{Zhou2019NatCommun,
  author  = {Zhou, Feng and Yan, Shuai and Zhou, Hailong and et al.},
  title   = {Field-Programmable Silicon Temporal Cloak},
  journal = {Nature Communications},
  year    = {2019},
  volume  = {10},
  pages   = {2726},
  doi     = {10.1038/s41467-019-10521-5}
}

@article{Wu2020TAP,
  author  = {Wu, Zhen and Grbic, Anthony},
  title   = {Serrodyne Frequency Translation Using Time-Modulated Metasurfaces},
  journal = {IEEE Transactions on Antennas and Propagation},
  year    = {2020},
  volume  = {68},
  number  = {3},
  pages   = {1599--1606},
  doi     = {10.1109/TAP.2019.2943712}
}

@article{Taravati2021PRApplied,
  author  = {Taravati, Sajjad and Eleftheriades, George V.},
  title   = {Pure and Linear Frequency-Conversion Temporal Metasurface},
  journal = {Physical Review Applied},
  year    = {2021},
  volume  = {15},
  number  = {6},
  pages   = {064011},
  doi     = {10.1103/PhysRevApplied.15.064011}
}

@article{Taravati2022ACSPhotonics,
  author  = {Taravati, Sajjad and Eleftheriades, George V.},
  title   = {Microwave Space-Time-Modulated Metasurfaces},
  journal = {ACS Photonics},
  year    = {2022},
  volume  = {9},
  number  = {2},
  pages   = {305--318},
  doi     = {10.1021/acsphotonics.1c01041}
}

@article{Peng2025ACSAEM,
  author  = {Peng, Shuang and Yu, Qian and Shen, Xiaoyue and Xie, Yating and Zhang, Han and Ma, Jie and Fu, Xiaojian and Wu, Junwei and Yang, Fei},
  title   = {Design and Implementation of a Millimeter-Wave Wireless Transceiver Based on a Waveguide-Fed Space-Time-Coding Metasurface},
  journal = {ACS Applied Electronic Materials},
  year    = {2025},
  volume  = {7},
  number  = {16},
  pages   = {7802--7810},
  doi     = {10.1021/acsaelm.5c01154}
}

@article{Galiffi2022AdvPhoton,
  author  = {Galiffi, Emanuele and Tirole, Romain and Yin, Shixiong and Li, Huanan and Vezzoli, Stefano and Huidobro, Paloma A. and Silveirinha, M{\'a}rio G. and Sapienza, Riccardo and Al{\`u}, Andrea and Pendry, J. B.},
  title   = {Photonics of time-varying media},
  journal = {Advanced Photonics},
  year    = {2022},
  volume  = {4},
  number  = {1},
  pages   = {014002},
  doi     = {10.1117/1.AP.4.1.014002}
}

@article{Drahushuk2016ACSNano,
  author  = {Drahushuk, Lee W. and Wang, Luda and Koenig, Steven P. and Bunch, J. Scott and Strano, Michael S.},
  title   = {Analysis of Time-Varying, Stochastic Gas Transport through Graphene Membranes},
  journal = {ACS Nano},
  year    = {2016},
  volume  = {10},
  number  = {1},
  pages   = {786--795},
  doi     = {10.1021/acsnano.5b05870}
}

@article{Salary2018PRB,
  author  = {Salary, Mohammad Mahdi and Jafar-Zanjani, Samad and Mosallaei, Hossein},
  title   = {Time-varying metamaterials based on graphene-wrapped microwires: Modeling and potential applications},
  journal = {Physical Review B},
  year    = {2018},
  volume  = {97},
  number  = {11},
  pages   = {115421},
  doi     = {10.1103/PhysRevB.97.115421}
}

@article{SensaleRodriguez2012NatComm,
  author  = {Sensale-Rodriguez, B. and Yan, R. and Kelly, M. and Fang, T. and Tahy, K. and Hwang, W. S. and Jena, D. and Liu, L. and Xing, H. G.},
  title   = {Broadband graphene terahertz modulators enabled by intraband transitions},
  journal = {Nature Communications},
  year    = {2012},
  volume  = {3},
  pages   = {780},
  doi     = {10.1038/ncomms1787}
}

@article{Sedeh2022TAP,
  author  = {Sedeh, H. Barati and Salary, Mohammad Mahdi and Mosallaei, Hossein},
  title   = {Active multiple access secure communication enabled by graphene-based time-modulated metasurfaces},
  journal = {IEEE Transactions on Antennas and Propagation},
  year    = {2022},
  volume  = {70},
  number  = {1},
  pages   = {664--679},
  doi     = {10.1109/TAP.2021.3119835}
}

@article{Ptitcyn2023LPR,
  author  = {Ptitcyn, G. and Lamprianidis, A. and Karamanos, T. and Asadchy, V. and others},
  title   = {Floquet--Mie theory for time-varying dispersive spheres},
  journal = {Laser \& Photonics Reviews},
  year    = {2023},
  volume  = {17},
  number  = {3},
  pages   = {2100683},
  doi     = {10.1002/lpor.202100683}
}

@article{Mostafa2022PRApplied,
  author  = {Mostafa, M. and D{\'\i}az-Rubio, A. and Mirmoosa, M. and Tretyakov, S. A.},
  title   = {Scattering from time-modulated metasurfaces: A general theoretical framework},
  journal = {Physical Review Applied},
  year    = {2022},
  volume  = {17},
  number  = {6},
  pages   = {064048},
  doi     = {10.1103/PhysRevApplied.17.064048}
}

@conference{Karamanos2024META,
  author    = {Karamanos, T. D. and Amanatiadis, S. and Kantartzis, N. and Lemoult, F.},
  title     = {Harmonic enhancement via a time-varying Fabry--Perot resonator},
  booktitle = {Proceedings of the 14th International Conference on Metamaterials, Photonic Crystals, and Plasmonics (META 2024)},
  address   = {Toyama, Japan},
  year      = {2024}
}

@article{hanson2008,
  title={Dyadic Green’s functions and guided surface waves for a surface conductivity model of graphene},
  author={Hanson, George W},
  journal={Journal of Applied Physics},
  volume={103},
  number={6},
  year={2008},
  publisher={AIP Publishing}
}

@article{christiansen2021inverse,
  title={Inverse design in photonics by topology optimization: tutorial},
  author={Christiansen, Rasmus E and Sigmund, Ole},
  journal={Journal of the Optical Society of America B},
  volume={38},
  number={2},
  pages={496--509},
  year={2021},
  publisher={Optical Society of America}
}

@book{OlverComplexAnalysis,
  author    = {Peter J. Olver},
  title     = {Complex Analysis and Conformal Mapping},
  year      = {2018},
  note      = {Lecture notes, University of Minnesota}
}

\newpage



  
  

\end{document}